 \documentclass[final,5p,times,twocolumn]{elsarticle}

\usepackage{fancybox}

\usepackage{fancyvrb}
\usepackage{color}
\usepackage{changepage} 

\usepackage[linesnumbered, ruled]{algorithm2e}
\usepackage{array}
\usepackage{framed}
\usepackage{xspace}
\usepackage{listings}
\usepackage[export]{adjustbox}

\newcommand{\rom}[1]{\uppercase\expandafter{\romannumeral #1\relax}}

\definecolor{gray50}{gray}{.5}
\definecolor{gray40}{gray}{.6}
\definecolor{gray30}{gray}{.7}
\definecolor{gray20}{gray}{.8}
\definecolor{gray10}{gray}{.9}
\definecolor{gray05}{gray}{.95}

\newlength\Linewidth
\def\findlength{\setlength\Linewidth\linewidth
\addtolength\Linewidth{-4\fboxrule}
\addtolength\Linewidth{-3\fboxsep}
}
\newenvironment{examplebox}{\par\begingroup
   \setlength{\fboxsep}{5pt}\findlength
   \setbox0=\vbox\bgroup\noindent
   \hsize=0.95\linewidth
   \begin{minipage}{0.95\linewidth}\normalsize}
    {\end{minipage}\egroup
    \textcolor{gray20}{\fboxsep1.5pt\fbox
     {\fboxsep5pt\colorbox{gray05}{\normalcolor\box0}}}
    \endgroup\par\noindent
    \normalcolor\ignorespacesafterend}

\newcounter{RQCounter}
\newcounter{RQACounter}

\newcommand{\RQ}[2]{%
\refstepcounter{RQCounter} \label{#1}
 \begin{center}	
  \begin{examplebox}
   \textbf{RQ\arabic{RQCounter}.}~#2
  \end{examplebox}	 
 \end{center}
}

\newcommand{\RS}[2]{%
\begin{framed}%
\filbreak
\textbf{Result {\ref{#1}}:~}{\emph {#2}}%
\end{framed}
}

\definecolor{javared}{rgb}{0.6,0,0} 
\definecolor{javagreen}{rgb}{0.25,0.5,0.35} 
\definecolor{javapurple}{rgb}{0.5,0,0.35} 
\definecolor{javadocblue}{rgb}{0.25,0.35,0.75} 

\lstdefinestyle{customc}{
  belowcaptionskip=\baselineskip,
  breaklines=true,
  xleftmargin=\parindent,
  language=java,
  showstringspaces=false,
  basicstyle=\scriptsize\ttfamily,
  keywordstyle=\bfseries\color{javapurple},
  commentstyle=\itshape\blue,
}

\usepackage{amssymb}
\usepackage{amsmath}
\usepackage{natbib}   

\usepackage[switch]{lineno}

\newcommand{\norm}[1]{\left\lVert#1\right\rVert}

\usepackage[colorlinks=true,linkcolor=blue]{hyperref}
\usepackage{kbordermatrix} 

\journal{The Journal of Systems and Software}

\begin{document}

\definecolor{greenx}{rgb}{0,0,0}
\definecolor{bluex}{rgb}{0,0,0}
\definecolor{redx}{rgb}{0,0,0}
\definecolor{redx2}{rgb}{0,0,0}
\definecolor{bluex2}{rgb}{0,0,0}
\definecolor{color10}{rgb}{0,0,0}
\definecolor{color11}{rgb}{0,0,0}
\definecolor{color12}{rgb}{0,0,0}
\definecolor{color13}{rgb}{0,0,0}

\begin{frontmatter}



\title{{Testing Multiple Linear Regression Systems with Metamorphic Testing}}


\author{Quang-Hung~Luu}
\author{Man~F.~Lau}
\author{Sebastian~P.H.~Ng\corref{cor}}
\author{Tsong~Yueh~Chen}
\address{Department of Computer Science and Software Engineering, Swinburne University of Technology, Hawthorn, Australia}

\cortext[cor]{Corresponding author}

\begin{abstract}
Regression is one of the most commonly used statistical techniques. However, testing regression systems is a great challenge because of the absence of test oracle. In this paper, we show that Metamorphic Testing is an effective approach to test multiple linear regression systems. In doing so, we identify intrinsic mathematical properties of linear regression, and then propose 11 Metamorphic Relations to be used for testing. Their effectiveness is examined using mutation analysis with a range of different regression programs. We further look at how the testing could be adopted in a more effective way. Our work is applicable to examine the reliability of predictive systems based on regression that has been widely used in economics, engineering and science, as well as of the regression calculation manipulated by statistical users.
\end{abstract}

\end{frontmatter}



\section{Introduction}

\noindent Being a cornerstone in statistics, regression is a fundamental prediction technique. Due to its simplicity and robustness, it has been widely used in numerous disciplines of science and engineering as well as other fields such as economics and business. Once a regression model is established, we are able to adopt it for predicting expected values within a certain level of confidence. 

Calculating regression coefficients is an essential statistical function of various software and systems. It has been integrated as part of both commercial software such as MATLAB (MathWorks), SPSS Statistics (IBM) and Excel (Microsoft) and open-source libraries including GNU Scientific Library (C/C++), {Scikit-learn, Scipy, StatsModels} (Python), Accord, MathNet (F\#), and GoLearn (Go). On the other hand, new systems may be built from scratch in some situations, such as the development of statistical libraries in new programming language. Besides, some programmers may need to modify the regression, for examples, transforming the variables \citep{lin2000, luu2015, wu2017, luu2018} or parallelizing the computations \citep{baboulin2006,dau2019}. In most cases, the calculation involves extensive computations. It is thus prone to programming bugs, and data mishandling. Blunders in computing regression by these software or systems would definitely cast doubts on the reliability and robustness of data and scientific analyses based on them.

Advancements in machine learning enable us to cope with the growth of big data, having the fastest rate over recent decades \citep{goodfellow2016}. Despite the rapid change, linear regression remains as the technique to be widely used due to two main reasons. First, it has been serving as a baseline for evaluating more sophisticated modelling systems for years \citep{weisberg2005,montgomery2012}. For example, simple linear regression is frequently adopted to assess the performance of various artificial neural networks (ANN) systems \citep{bibault2018, ahmad2019, mondal2020, ayoubloo2011}. Second, the ANN equations are reducible to the linear regression form, being referred to as deep linear network \citep{saxe2014, lathuiliere2019}. Recently, the deep linear network has received lots of attention as it helps to better understand the properties and reliability of general ANN \citep{lathuiliere2019, lee2019, bah2020}. Reliability of many systems therefore somehow relies on the correctness of linear regression calculation.


Traditionally, the testing of regression system is difficult because we do not have a test oracle in general. For example, if we have some data points, say $(2, 4), (7, 6)$ and $(9, 10)$, and want to find a line that has the predicted values ``closest'' to these points. (In case of higher dimensions, the plane or hyperplane is used instead of line.) When the regression program returns the line as $y = x + 3$, we have no simple way to tell whether or not this line is the best fitted line. Testers may find the regression line manually by solving the required equations constructed from these data points. However, this is not viable and is very tedious.


Testing such programs becomes even more difficult in reality because we often never know the {\it true line} due to round-off errors \citep{higham2002,solomon2015}. In other words, we are not able to determine whether or not the computed value from a regression algorithm is the exact estimator that should be returned by the algorithm. In fact, previous studies \citep{higham2002,gratton2013,marland2018} pointed out that solving the linear regression may cause a large accumulation of round-off and truncation errors because achieving the regression solutions requires intensive matrix calculations. When the true value is unavailable, there is no direct way to validate the regression systems.

Incorrect linear regression computation may lead to severe consequences. In many cases, even a marginal difference in estimating coefficients of regression may cause a huge impact. For instance, while a previous study estimated the global mean sea level has been rising at a rate of $1.7$ mm per year since the beginning of this century \citep{church2011}, another work suggested that the rate should be $1.2$ mm per year \mbox{\citep{hay2015}}. Due to the acceleration of global warming, the two regression coefficients after 50 years will lead to a gap larger than 2.5 cm in the predicted sea level that miscounted multi billion dollars per year in flood damage cost, according to a recent research \citep{jevrejeva2018}. Given that fact, while round-off error was attributed to the US Army's Patriot Missile Defense System failure in shooting down the SCUD missile causing death of 28 American soldiers \citep{zhivich2009}, the rounding is also known to deteriorate the accuracy of regression \citep{higham2002}. Indeed, a faulty regression program may cause damages more severe than the round-off error. Validating the accuracy of regression is therefore not solely a computational issue but also has real-life impacts. Considering its very diversified range of applications, testing the correctness of regression systems is certainly of great importance.

In view of such problem, we should find an appropriate approach to tackle the testing of linear regression-based systems. The software testing technique of Metamorphic Testing (MT), does allow us to alleviate the absence of oracle, and to generate test sets automatically \citep{segura2016,chen2017}. It has been applied successfully in testing various software systems, such as numerical analysis \citep{chan1998}, partial differential equations \citep{chen2002} search engines \citep{zhou2012,zhou2016}, Web Application Programming Interfaces \citep{segura2018}, feature models \citep{segura2010, segura2011}, scientific software \citep{ding2016,lin2018}, programming language compilers \citep{le2014}, graphics shader compilers \citep{donaldson2017}, context-sensitive middleware-based applications \citep{chan2005}, object detection \citep{wang2019}, cybersecurity \citep{chen2016,mai2020}. 

Recently, MT has been applied efficiently to validate various machine learning and artificial intelligence systems \citep{xie2011,xie2018,segura2018b}. A testing tool named DeepTest was developed to detect the erroneous responses of autonomous car-driving systems backed by deep neural networks (DNN) \citep{tian2018}. It helped revealing thousands of dangerous situations under different weather conditions affecting the functioning of DNN-based programs that were submitted to the Udacity self-driving challenge. Another example in testing a real-life driver-less car system named Apollo of Baidu \citep{zhou2019} pointed out that the random noises added will significantly alter the system's responses, potentially leading to fatal crashes. In fact, MT has been evolving beyond the domain of conventional testing. It is extended to validation \citep{lin2018,zhou2019}, program understanding \citep{zhou2018} and quality assessment \citep{pullum2012}. MT was also combined with ``statistical hypothesis test'' to deduce the likelihood of a system's faultiness \citep{guderlei2007}. Despite of its many successes with complex systems, it remains unclear to what extent that MT is applicable to test linear regression-based systems.

The aim of this paper is to propose the use of MT to test multiple linear regression systems. We focus on testing the implementation of linear regression algorithm for three reasons. First of all, the algorithm is the core component of these regression systems. While testing a specific system may require a particular set of features and properties specific to that system, the technique for testing such system may not be applicable to test another system. For example, testing a C++ program may involve an explicit declaration of variables, while it is not required in testing a Python or Ruby program, making dissimilar outcomes in assessing the bugs associated with initializing variables in different systems. Second, we may inspire interested readers to come up with new intrinsic properties, or extend them for non-linear regression-based systems. Third, it has been reported that many statistical users have mistakenly manipulated the regression calculation, and thus might derive incorrect predictive models. For example, many users came up with erroneous results while applying linear regression in Python \citep{bug-python1,bug-python2,bug-python3}, Matlab \citep{bug-matlab2} or R \citep{bug-r1,bug-r2} software, and it turned out that they manipulated the regression wrongly by mishandling the intercept. By considering our different ways to check the implementation of regression algorithm, they may be able to test their own manipulation process, and thus improve the reliability of their works.  

For MT to work on the regression systems, testers or users need to know some intrinsic properties of the regression. These properties should be supported by formal ``mathematical proofs'' to avoid misuses. For example, there has been some reports about users adopting some properties that are not necessary properties of the algorithms to be implemented \citep{xie2011}. This will cause misinterpretations of the results. One of the strengths of this paper is that we prove all these intrinsic properties of the regression systems. Contributions of this paper are summarized as follows:
\begin{itemize}
  \item We derive a set of mathematical properties of estimator of multiple linear regression related to the addition of data points, the rescaling of inputs, the shifting of variables, the reordering of data, and the rotation of independent variables.
  \item We then develop 11 Metamorphic Relation (MRs) which are grouped in 6 categories to support the testing of related regression systems.
  \item The technique of mutation analysis is applied on a wide range of regression programs to examine the performance of these MRs. 
   \item We further examine how the testing of regression could be carried out in a more effective way.
\end{itemize}

The structure of this paper is organised as follows. After the introduction, we start with recalling basic concepts of linear regression and MT. In Section 3, we present selected properties of estimator, and associated MRs that will later be used for verifying the regression estimates. Section 4 is on experimentation, presenting the mutation set-up, the generation of the input datasets, and the assessment criteria. The effectiveness of MRs is reported and discussed in Section 5, before a comparison between MT and random testing is presented in Section 6. We describe the threats to validity of our experiments in Section 7. Finally, the paper is enclosed by concluding remarks in Section 8.


\section{Preliminaries}

\subsection{Multiple linear regression}

Regression is a model to describe the behavior of a certain random variable of interest. This variable may be the price of stocks in the financial market, the growth of a biological species, or the detection opportunity of the gravitational wave. It is referred to as the dependent variable, and denoted with $y$. Information of the dependent variable is provided on the basis of predictor or explanatory variables; such as time for the price of stocks, food for the species to grow or the number of observations for the detection of wave. These terms are usually called the {\it independent variables}, and denoted with $x_0, x_1, \ldots ,x_d$. The regression model is to relate dependent variable to given independent variables by means of a function $f$

\begin{equation}
 y \approx f(x_0, x_1, \ldots ,x_d)
\end{equation}

Given the form of function $f$, the regression model is said to be determined when the unknown parameter, being referred to as the {\it estimator} $\beta$, is obtained. The optimal estimator $\hat\beta$ is chosen such that the modelled dependent variable $\hat{y}$ becomes closest to the true dependent variable $y$. In multiple linear regression model, the variable $\hat{y}$ has a linear relationship with respect to the components of optimized estimator $\hat\beta$ by means of:

\begin{equation}\label{equation:nointercept}
	\hat y = x_0 \hat\beta_0 
				+ x_1 \hat\beta_1
				+ \ldots
				+ x_d \hat\beta_d
\end{equation}
where $\hat\beta = (\hat\beta_0, \hat\beta_1, \ldots, \hat\beta_d)^T$ in which the upper symbol $T$ represents the transpose of the relevant matrix. Each individual variable $x_k$ ($k$ = 0, 1, \ldots, $d$) can be, for instance, a trend rate, an oscillatory factor or a function in the multiple linear regression \citep{luu2015,luu2018}, but it must be independent of all other variables $x_i (\forall i\neq k)$. The intercept form is adopted if we set $x_0$ as a constant, say $x_0=1$,  in contrast with other input variables $x_1, x_2, \ldots, x_d$ that are fed with data. Then we have
\begin{equation}\label{equation:intercept}
	\hat{y} = \hat\beta_0 
				+ x_1 \hat\beta_1
				+ \ldots
				+ x_d \hat\beta_d
\end{equation}

Assume we have $n$ data points for the model, and denote ${\bf x}_0, {\bf x}_1, \ldots, {\bf x}_d$ and ${\bf y}$ (all $\in {\rm I\!R}^{n}$) as corresponding vectors of data for the variables $x_0, x_1, \ldots, x_d$ and $y$, {and $\hat{\bf y}$ {($\in {\rm I\!R}^{n}$)} is the vector of modelled dependent variable $\hat{y}$ .
Their matrices are expressed as 
}
\begin{equation}\label{equation:denote}
\begin{aligned}
  {\bf X} =
  \begin{bmatrix}
    {\bf x}_{0} \\
    {\bf x}_{1} \\
    \ldots\\
    {\bf x}_d \\
  \end{bmatrix}^T &=   
  \begin{bmatrix}
    x_{1,0} & x_{2,0} & \ldots  & x_{n,0} \\
    x_{1,1} & x_{2,1} & \ldots  & x_{n,1} \\
    \ldots & \ldots & \ldots  & \ldots \\
    x_{1,d} & x_{2,d} & \ldots  & x_{n,d}    
  \end{bmatrix} 
  ;\\
  {\bf y} &=
  \begin{bmatrix}
    {y}_{1} \\
    {y}_{2} \\
    \ldots\\
    {y}_{n}
  \end{bmatrix}
  {
  ;
  {\bf\hat y} =
  \begin{bmatrix}
    \hat{y}_{1} \\
    \hat{y}_{2} \\
    \ldots\\
    \hat{y}_{n}
  \end{bmatrix}
  }
\end{aligned}
\end{equation}
The linear regression equation to derive $\hat{\bf y}$ is
\begin{equation}
	\hat{\bf y} = {\bf X}^T \hat\beta
\end{equation}

To establish the model, a cost function (also named as objective function) is adopted to quantify the estimator $\hat\beta$ based on a certain metrics. The idea is to minimize the difference between the true ${\bf y}$ and the modelled $\hat{\bf y}$, which is referred to as the residual (or error). In general, the cost function consists of the residual, the estimator and different

forms of parameters and weights. Some regression methods may add a penalty quantity associated with $\hat\beta$, or take into account the standard deviation. In the ordinary least square (OLS) regression, the cost function is simply defined as the total of square of residuals. The optimized estimator is determined by the minimization ($\mathrm{arg~min}$) of this cost function using the square of Euclidean norm ($\norm{.}_2$), that is
\begin{equation}\label{equation:beta2}
	\hat\beta = \underset{\beta}{\mathrm{arg~min}} \norm{ {\bf y} - \hat{\bf y}  } _2^2
\end{equation}
There are different ways to derive the solution of this equation such as derivatives of sum of square errors, maximum likelihood, projection, generalized method of moments. When the matrix ${\bf X}{\bf X}^T$ has the full rank, i.e. all its rows and columns are linearly independent, we have the unique solution derived from the {\it normal equation}:
\begin{equation}\label{equation:beta}
	\hat\beta = ({\bf X}{\bf X}^T)^{-1} {\bf X}{\bf y}
\end{equation}
The upper numeric subscript ($-1$) represents the inverse of a related matrix. Once the model is established after the derivation of the estimator $\hat\beta$, we may predict the change of dependent variable $\hat{y}$ under varying conditions of $\{{x}_0,{x}_1,\ldots,{x}_d\}$. In the prediction system based on linear regression, the core model is trained with a set of input data to obtain the optimized estimator ($\hat{\beta}$), before it can be used to predict. A more comprehensive description about linear regression can be found in \citep{weisberg2005,montgomery2012}.

\subsection{Metamorphic testing}


Metamorphic Testing (MT) has been developed to alleviate the \emph{test oracle} problem, which is referred to as the situations where test results of a program are impossible or extremely difficult to be validated. Consider an example of a program $P$ that implements a particular algorithm $A$ to find the shortest path from one node to another in a graph. Assume that $G$ is a graph having 100 nodes, and that, on average, each node has about 20 edges. In general, given the graph $G$ and two nodes $x$ and $y$ in $G$, it is very time-consuming to verify that the program's actual output $P(G,x,y)$ is really a shortest path in $G$ from $x$ to $y$ because a manual process may involve validating against $100!$ possible paths connecting x and y.

The core of MT is the concept of Metamorphic Relations (MRs), which are derived from properties of the targeted algorithm to be implemented. If the program does not uphold these properties, we can conclude that the program has errors. Using the shortest path example mentioned earlier, assume further that $x$, $y$ and $z$ are three different nodes in $G$. Based on the domain knowledge of the shortest path in a graph, one can derive the metamorphic relation -- ``if $z$ is in the shortest path in $G$ from $x$ to $y$, the length of the shortest path in $G$ from $x$ to $y$ is equal to the sum of the length of the shortest path in $G$ from $x$ to $z$ and that from $z$ to $y$''. Since the program $P$ is implementing a particular ``shortest path'' algorithm, one can expect that $P$ upholds this metamorphic relationship in the sense that ``If $z$ is in the actual output $P(G,x,y)$, the length of $P(G,x,y)$ is equal to the sum of the lengths of $P(G,x,z)$ and $P(G,z,y)$.
The original test case $(G,x,y)$ is considered as a \emph{source test case} in MT.
The two test cases $(G,x,z)$ and $(G,z,y)$ are referred to as the \emph{follow-up test cases} in MT because these test cases are follow-ups of the source test case.
In this example, the MR involves one source test case and two follow-up test cases, and the follow-up test cases may depend on the actual output of the program with the source test case as input.
In general, an MR can involve multiple source test cases and multiple follow-up test cases.
Following is the formal definitions of MR and other concepts used in MT \citep{chen2017}.

\vspace{0.25cm}
\noindent\textbf{Definition 1:} 
\textit{Let $g$ be a target function or algorithm to be implemented.
A \textbf{Metamorphic Relation (MR)} is a necessary property of $g$ over a sequence having two or more inputs $\operatorname{<}I_1, I_2, \ldots, I_m\operatorname{>}$, where $m\geq 2$, and the sequence of corresponding outputs $\operatorname{<} g(I_1), g(I_2), \ldots, g(I_m)\operatorname{>}$. The relation is denoted by $\mathcal{R} \subseteq X^{m} \times Y^{m}$, where $\subseteq$ is the subset relation, and $X^{m}$ and $Y^{m}$ are the Cartesian products of $m$ input and $m$ output spaces, respectively. } 
We may simply adopt $\mathcal{R} \left(I_1, I_2, \ldots, I_m, g(I_1), g(I_2), \ldots,g(I_m)\right)$
to represent $\operatorname{<}I_1, I_2, \ldots, I_m, g(I_1), g(I_2), \ldots, g(I_m)\operatorname{>} \in \mathcal{R}$.

\vspace{0.25cm}
\noindent\textbf{Definition 2:} 
\textit{Consider an MR $\mathcal{R} \left(I_1, I_2, \ldots, I_m, g(I_1), g(I_2), \ldots, g(I_m)\right)$. 
Suppose that each $I_j$ ($j=k+1,k+2,\ldots,m$) is constructed based on $\operatorname{<}I_1, I_2, \ldots, I_k, g(I_1), g(I_2), \ldots, g(I_k)\operatorname{>}$ according to $\mathcal{R}$. For any $i=1,2,\ldots,k$, $I_i$ is referred to as a \textbf{source input}. For any $j=k+1,k+2,\ldots,m$, $I_j$ is referred to as a \textbf{follow-up input}. That is, for a given $\mathcal{R}$, if all source inputs $I_i$ ($i=1,2,\ldots,k$) are specified, then the follow-up inputs $I_j$ ($j=k+1,k+2,\ldots,m$) can be constructed based on the source inputs and, if necessary, their corresponding outputs. The sequence of inputs $\operatorname{<}I_1, I_2, \ldots, I_m\operatorname{>}$ is referred to as a metamorphic test group (MTG) of inputs for the MR.
} 

\vspace{0.25cm}
\noindent\textbf{Process for Metamorphic Testing (MT).}
\textit{Let $\mathcal{P}$ be an implementation of a target function $g$. For an MR $\mathcal{R}$, suppose that we have $\mathcal{R} \left(I_1, I_2, \ldots, I_m, g(I_1), g(I_2), \ldots, g(I_m)\right)$. \textbf{Metamorphic Testing (MT)} based on this MR for $\mathcal{P}$ involves the following steps:
} 

\textit{
(1) Define $\mathcal{R}'$ from $\mathcal{R}$ by replacing $g$ by $\mathcal{P}$ 
} 

\textit{
(2) Given a sequence of source test cases $\operatorname{<}I_1, I_2, \ldots, I_k\operatorname{>}$. After execution, their respective \textbf{source outputs} are given by $\operatorname{<}\mathcal{P}(I_1), \mathcal{P}(I_2), \ldots, \mathcal{P}(I_k)\operatorname{>}$. Construct and execute a sequence of follow-up test cases $\operatorname{<}I_{k+1}, I_{k+1}, \ldots, I_m\operatorname{>}$ with reference to $\mathcal{R}'$, and obtain their corresponding \textbf{follow-up outputs} $\operatorname{<}\mathcal{P}(I_{k+1}), \mathcal{P}(I_{k+1}), \ldots, \mathcal{P}(I_m)\operatorname{>}$.
} 

\textit{
(3) Compare the executed results against the expectation given by $\mathcal{R}'$. If $\mathcal{R}'$ is not satisfied, then $\mathcal{P}$ is determined to be faulty by this MR. 
} 

\vspace{0.25cm}

In other words, to apply MT to test a program $\mathcal{P}$ using a given MR, we first need to generate the relevant source test cases. We then execute the program $\mathcal{P}$ with these source test cases to obtain their respective source outputs. Based on the given MR, we can generate the relevant follow-up test cases. After that, the program $\mathcal{P}$ is executed with these follow-up test cases to obtain the respective follow-up outputs. Finally, we can check whether the given MR is satisfied by the source test cases, the source outputs, the follow-up test cases, and the follow-up outputs. If the MR is violated, it means that the program under test $\mathcal{P}$ is faulty. This entire process of MT can be automated.

\vspace{0.25cm}
\section{Metaphoric relations of estimator}
In this section, we propose 11 MRs that can be used to test linear regression based systems. As mentioned earlier, they are derived from the properties of the targeted algorithm to be implemented. It is important to discuss the intuition of each property, understand the properties and explain how MRs could be derived from the relevant properties. Interested readers may find the mathematical proofs of these properties in Appendix~A. For ease of discussion and illustration, we will use 2-dimensional examples, if needed. Last, but not least, all properties discussed in this section involve the optimized estimator $\hat{\beta}$.


\vspace{0.25cm}
\subsection{Properties of estimator}\label{section:Property}
\subsubsection{Property 1. Inserting new data}

This property is about inserting an additional data point to the original set of data points for linear regression. Proposition~1 shows that, after a new data point is added to the original set of data points, the new linear regression line obtained from the new set of data points can be derived from a relationship with the original regression line and the new data point.

\vspace{0.25cm}
\noindent\textbf{Proposition 1. }

\textit{Suppose that the dependent variable $y$ is related to a linear relationship with independent variables $x_0, x_1, \ldots, x_d$ with the estimator $\hat\beta$ derived from the least square fitting. 
Let ${\bf y}$ and {${\bf x}_0, {\bf x}_1, \ldots, {\bf x}_d$} (all $\in {\rm I\!R}^{n}$) denote the vectors of data for the variables $y$ and $x_0, x_1, \ldots, x_d$, respectively, whose matrices are expressed in Equation~(\ref{equation:denote}).
And let $x^* \in {\rm I\!R}^{d+1}$ and ${y^*} \in {\rm I\!R}$ be a data point being added into the original data set. The linear estimator $\hat\beta^*$ obtained from the new data set $ {\bf X}^* = \begin{bmatrix} {\bf X} & x^* \end{bmatrix}  \in {\rm I\!R}^{(d+1)\times (n+1)}$ and $ {\bf y}^* = \begin{bmatrix} {\bf y} & {y^*} \end{bmatrix} \in {\rm I\!R}^{n+1}$ can be derived from the following equation
\begin{align}\label{equation:betachange}
    \hat\beta^* 
    = \hat\beta+ G ({y^*}-x^{*T}\hat\beta)
\end{align}
where $G$ is the vector of size $d+1$ defined by $\bf{X}$ and $x^*$ as follows
\begin{equation}
    G = \frac
        {({\bf X}{\bf X}^T)^{-1} x^*}
        {1+x^{*T}({\bf X}{\bf X}^T)^{-1} x^*}.
\end{equation}
} 

We now derive an important property that forms the foundation of the first two metamorphic relations, namely MR 1.1 and MR 1.2 in Section~\ref{section:MR}.

\vspace{0.25cm}
\noindent\textit{\textbf{Corollary~1. }Same notation as in Proposition~1. 
If $y^* = x^{*T}\hat\beta$, then} $\hat\beta^* = \hat\beta$.

\noindent\textit{Proof: }
Suppose $y^* = x^{*T} \hat\beta$. It is straightforward from Equation~(\ref{equation:betachange}) that $\hat\beta^* = \hat\beta$.

In other words, if the new data point is generated from the model derived by the original data set, we do expect that the new model obtained from the new data set will be the same as the original model.
For example, suppose we have 3 data points $(x = 1, y = 3), (3, 7), (5, 11)$ in the data set for linear regression and that the linear regression system {returns} the line $y = 2x + 1$.
If we use this line equation to generate a new data point, say (7, 15), and feed all 4 (original 3 plus this new) data points to the linear regression system, we expect that the new regression line obtained will be the same as the original one.

\vspace{0.25cm}
\noindent\textit{\textbf{Corollary~2. }Same notation as in Proposition~1. If
\begin{align}
\overline{y} = \frac{1}{n}\sum_{i=1}^n y\\
\overline{x_{j}} = \frac{1}{n}\sum_{i=1}^n x_{i,j}
\end{align}
for $j=1,2,\ldots,d$, then $\hat\beta^* = \hat\beta$ for the regression with intercept, being explicitly expressed in Equation~(\ref{equation:intercept}).}

\noindent\textit{Proof: }
The summation of both sides of Equation~(\ref{equation:intercept}) gives us
\begin{align}
\sum_{i=1}^n \hat{y} = \sum_{j=1}^d \hat\beta_j \sum_{i=1}^n x_{i,j} 
\end{align}
The linear regression with intercept is unbiased, that is
\begin{align}
\frac{1}{n}\sum_{i=1}^n \hat{y} = \frac{1}{n}\sum_{i=1}^n y 
\end{align}
As a result, from the definitions of $\overline{x}$ and $\overline{y}$, we obtain
\begin{align}
\overline{y} =\sum_{j=1}^d \hat\beta_j \overline{x_{j}}
\end{align}
This equation can be rewritten in form of vector as $\hat y = x^{*T}\hat\beta$. {It follows immediately after Corollary~1 that} $\hat\beta^* = \hat\beta$.


\subsubsection{Property 2. Scaling data}

The second property is about the scaling of certain coordinates of data points. Proposition~2 shows that, if certain coordinates of data points along a certain axis are scaled by a given factor, the slope of regression line with respect to this axis will be scaled proportionally by the same factor.

\vspace{0.25cm}
\noindent\textbf{Proposition 2. }
\textit{Suppose that the values ${\bf y}^*$ of the dependent variable are scaled by a factor of $a$ with respect to the original values ${\bf y}$, and the values of an independent variables ${\bf x}^*_k$ being factorized by a factor of $b$ with respect to the original values ${\bf x}_k$, that is 
\begin{align}
{\bf y}^* &= a~ {\bf y}\\
{\bf x}^*_k &= b ~{\bf x}_k 
\end{align}
where the constants $a$ and $b$ are non-zero real numbers ($a, b \in {\rm I\!R}\backslash \{ 0 \}$). The new estimator $\hat\beta^*$ can be computed from the original estimator $\hat\beta$ as follows
\begin{equation}
\begin{aligned}\label{equation:beta-scaled}
  \hat{\beta}^* = 
  \begin{bmatrix}
    a\hat\beta_{0} \\
    a\hat\beta_{1} \\
    \ldots\\
    a\hat\beta_{k-1} \\
    \frac{a}{b}\hat\beta_{k} \\
    a\hat\beta_{k+1} \\
    \ldots\\
    a\hat\beta_d
  \end{bmatrix}
\end{aligned}
\end{equation}
}

Based on this proposition, we can further derive two properties of mirroring the components of the data points, and two properties of scaling the components of the data points. These properties are the basis of four MRs, namely MR 2.1, MR 2.2, MR 3.1 and MR 3.2 in Section~\ref{section:MR}. The proofs of these properties are straightforward from Equation~ (\ref{equation:beta-scaled}). We will leave them to the readers.

\vspace{0.25cm}
\noindent\textit{
\textbf{Corollary~3. }Same notation as in Proposition~2. If $a = -1$ and $b = 1$, then $\hat\beta^* = - \hat\beta$.
}

That is, if we reflect the sign of the dependent variable, we expect that the sign of estimator will also be reflected. This is the basis of our MR 2.1.

\vspace{0.25cm}
\noindent\textit{\textbf{Corollary~4. }Same notation as in Proposition~2. If $a = 1$ and $b = -1$, then $\hat\beta^*$ is given by the following equation
\begin{equation}
\begin{aligned}
  \hat{\beta}^* = 
  \begin{bmatrix}
    \hat\beta_{0} \\
    \hat\beta_{1} \\
    \ldots\\
    \hat\beta_{k-1} \\
    - \hat\beta_{k} \\
    \hat\beta_{k+1} \\
    \ldots\\
    \hat\beta_d
  \end{bmatrix}
\end{aligned}
\end{equation}
}

More simply, if we only reflect a particular independent variable, only the component of estimator related to this independent variable is reflected. This is the basis of our MR 2.2.

\vspace{0.25cm}
\noindent\textit{\textbf{Corollary~5. }Same notation as in Proposition~2. If $a > 0$ and $b=1$, then $\hat\beta^* = a \hat\beta$.
}

Intuitively, if we scale the dependent variable by a factor of $a~(>0)$, the estimator will also be scaled by the same factor. This is the basis of our MR 3.1.

\vspace{0.25cm}
\noindent\textit{\textbf{Corollary~6. }Same notation as in Proposition~2. If $a=1$ and $b~(>0)$, then $\hat\beta^*$ is given by the following equation
\begin{equation}
\begin{aligned}
  \hat{\beta}^* = 
  \begin{bmatrix}
    \hat\beta_{0} \\
    \hat\beta_{1} \\
    \ldots\\
    \hat\beta_{k-1} \\
    \frac{1}{b} \hat\beta_{k} \\
    \hat\beta_{k+1} \\
    \ldots\\
    \hat\beta_d
  \end{bmatrix}
\end{aligned}
\end{equation}
}

That is, if we only scale a particular independent variable by a factor of $b~(>0)$, only the component of the estimator related to this independent variable is scaled reciprocally. This is the basis of our MR 3.2.

\subsubsection{Property 3. Shifting data}

The third property is about the shifting of certain components of the data in the original data set. Proposition~3 shows that, if the coordinates of data points along a certain axis are shifted by a given distance, the projection of the regression line {with intercept} is also shifted by a proportional distance.

\vspace{0.25cm}
\noindent\textbf{Proposition 3. }
\textit{Suppose that the values ${\bf y}^*$ of the dependent variable are shifted by a distance of $a$ with respect to the original values ${\bf y}$, and the values of an independent variables ${\bf x}^*_k$ being shifted by a distance of $b$ with respect to the original values ${\bf x}_k$, that is
\begin{align}
{\bf y}^* &= {\bf y} + a~{{\bf 1}^{n}}\label{equation:shifted-y}\\
{\bf x}^*_k &= {\bf x}_k + b ~{{\bf 1}^{n}}\label{equation:shifted-x}
\end{align}
where $a$ and $b$ are real constants ($a, b \in {\rm I\!R}$). In the linear regression with intercept, the new estimator $\hat\beta^*$  can be determined from the original estimator as follows
\begin{equation} \label{equation:beta-shifted}
\begin{aligned}
  \hat{\beta}^* = 
  \begin{bmatrix}
    \hat\beta_{0} - b \hat\beta_{k} + a \\
    \hat\beta_{1} \\
    \hat\beta_{2} \\
    \ldots\\
    \hat\beta_d
  \end{bmatrix}
\end{aligned}
\end{equation}
} 

Based on this proposition, we can further derive two properties of shifting the components of the data points. These properties are the basis of two MRs, namely MR 4.1 and MR 4.2 in Section~\ref{section:MR}.
The proofs of these properties are straightforward from Equation~(\ref{equation:beta-shifted}). 

\vspace{0.25cm}
\noindent\textit{\textbf{Corollary~7. }Same notation as in Proposition~3. If {$a > 0$} and $b=0$, then $\hat\beta^*$ is given by the following equation
\begin{equation}
\begin{aligned}
  \hat{\beta}^* = 
  \begin{bmatrix}
    \hat\beta_{0} + a \\
    \hat\beta_{1} \\
    \hat\beta_{2} \\
    \ldots\\
    \hat\beta_d
  \end{bmatrix}
\end{aligned}
\end{equation}
}

In other words, if we add up the dependent variable by a value of $a~(>0)$, the intercept component of estimator will also be increased by the same value. This is the basis of our MR 4.1.

\vspace{0.25cm}
\noindent\textit{\textbf{Corollary~8. }Same notation as in Proposition~3. If $a=0$ and $b > 0$, then $\hat\beta^*$ is given by the following equation
\begin{equation}
\begin{aligned}
  \hat{\beta}^* = 
  \begin{bmatrix}
    \hat\beta_{0} - b \hat\beta_{k} \\
    \hat\beta_{1} \\
    \hat\beta_{2} \\
    \ldots\\
    \hat\beta_d
  \end{bmatrix}
\end{aligned}
\end{equation}
}

That is, if we add up a particular independent variable by a value of $b~(>0)$, the intercept component of estimator will also be decreased by an amount proportional to both this value and the component of estimator related to this variable. This is the basis of our MR 4.2.


\subsubsection{Property 4. Permuting data}

The fourth property is about the permutation of certain components or certain samples of the data. Proposition~4  shows that, if the coordinates of the data points along certain axes are permuted, the corresponding projections of the regression line along these axes are permuted in the same order. Moreover, the proposition shows that, permuting the data points does not change the regression line.

\vspace{0.25cm}
\noindent\textbf{Proposition 4. }
\textit{Suppose that the samples ${\bf y}^*$ of the dependent variable are permuted by the function $\sigma_{s}$ with respect to the original variable ${\bf y}$; and the samples ${\bf x}^*_0,{\bf x}^*_1,\ldots,{\bf x}^*_d$ of the independent variables are permuted by both functions $\sigma_{s}$ and $\sigma_{v}$ with respect to the original values $\{{\bf x}_0,{\bf x}_1,\ldots,{\bf x}_d\}$, such that
\begin{align}
	{\bf y}^* &= \sigma_s({\bf y})\label{equation:permuted-sample}\\
	{\bf x}^*_k &= \sigma_s({\bf x}_{\sigma_v(k)})\label{equation:permuted-both}
\end{align}
where $\sigma_{v}$ is a permutation (bijective function) from set $\{{\bf x}_0,{\bf x}_1,\ldots,{\bf x}_d\}$ to $\{{\bf x}^*_0,{\bf x}^*_1,\ldots,{\bf x}^*_d\}$; whilst $\sigma_{s}$ is the corresponding bijective renumbering of the set of sample index $\{1, 2,\ldots,n\}$. 
The new estimator $\hat\beta^*$ can be determined from the original estimator as follows
\begin{equation} \label{equation:beta-permuted}
\begin{aligned}
  \hat{\beta}^* = 
  \begin{bmatrix}
    \hat\beta_{\sigma_v(0)} \\
    \hat\beta_{\sigma_v(1)} \\
    \ldots\\
    \hat\beta_{\sigma_v(d)}
  \end{bmatrix}
\end{aligned}
\end{equation}
} 

Based on this proposition, we can further derive two properties of permuting the data points. These properties are the basis of two MRs, namely MR 5.1 and MR 5.2 in Section~\ref{section:MR}. The proofs of these properties are straightforward from Equation~(\ref{equation:beta-permuted}). 

\vspace{0.25cm}
\noindent\textit{\textbf{Corollary~9. }Same notation as in Proposition~4. If $\sigma_v$ is the identity function (i.e., $\sigma_v(k)=k$) and $\sigma_s$ is a bijective function, then $\hat{\beta}^*=\hat{\beta}$.
}

{Put it differently}, if we permute the data points { only without permuting the dependent and independent variables}, the estimator will {remain} unchanged.
This is the basis of our MR 5.1.

\vspace{0.25cm}
\noindent\textit{\textbf{Corollary~10. }Same notation as in Proposition~4.  If $\sigma_v$ is a permutation that only swaps ${\bf x}_p$ and ${\bf x}_q$ where $0 \leq p, q \leq d$ and $p \neq q$, while $\sigma_s$ is the identity function, then $\hat{\beta}^*$ is given by the following equation
\begin{equation}
 \hat\beta_j^* = 
     \begin{cases}
        \hat\beta_{q} &\quad\text{for } j = p\\     
        \hat\beta_{p} &\quad\text{for } j = q\\     
        \hat\beta_{j} &\quad\text{otherwise}
     \end{cases}
\end{equation}
}

That is, if we swap two independent variables and their relevant values in the data points, the corresponding components of the estimator will be swapped accordingly. This is the basis of our MR 5.2.

\subsubsection{Property 5. Rotating data}

The fifth property is about the rotation of coordinate system. Proposition~5 shows that, if the axes related to the independent variables of the data points are rotated by a given rotation angle, the corresponding projections of the regression line are rotated by the same angle.

\vspace{0.25cm}
\noindent\textbf{Proposition 5. }
\textit{
Suppose that the values ${\bf y}^*$ of the dependent variable is kept unchanged, while the components ${\bf x}^*_0,{\bf x}^*_1,\ldots,{\bf x}^*_d$ of the independent variables are rotated by the matrix ${\bf R}$ of size $(d+1)\times (d+1)$. In other words, ${\bf R}$ rotates the matrix ${\bf X}^*$ with respect to the original matrix ${\bf X}$, that is
\begin{equation}\label{equation:rotated-beta-a}
	{\bf X}^* = {\bf R} {\bf X} 
\end{equation} 
Then the new estimator $\hat\beta^*$ can be determined from the original estimator as follows
\begin{equation}\label{equation:rotation2}
	\hat\beta^* = {\bf R}\hat\beta
\end{equation} 
}

Based on this proposition, we can derive the property of rotating the data points. This property is the basis of the MR 6 in Section~\ref{section:MR}. The proof of this property is straightforward from Equation~(\ref{equation:rotation2}). 

\vspace{0.25cm}
\noindent\textit{\textbf{Corollary~11. }Same notation as in Proposition~5. 
Suppose two components ${\bf x}_p$ and ${\bf x}_q$ with $1\leq p, q\leq d, p \neq q$ are rotated by an angle $\theta$ in the counter-clockwise direction, that is
\begin{equation}
\begin{cases}
   {\bf x}_p^* &= {\bf x}_p \cos\theta - {\bf x}_q \sin\theta \\
   {\bf x}_q^* &= {\bf x}_p \sin\theta + {\bf x}_q \cos\theta 
  \end{cases}
\end{equation}
in other words, the rotation matrix has the form 
\begin{equation}
\begin{aligned}
   &{\bf R}_{pq} = &\kbordermatrix{
      &  &  &  &  p & & q & \\
      & 1 & 0 & \ldots & 0 & \ldots & 0  & \ldots  & 0 \\
      & 0 & 1 & \ldots & 0 & \ldots & 0  & \ldots  & 0 \\
      & \ldots & \ldots & \ldots & \ldots &   \ldots & \ldots & \ldots & \ldots \\
    p & 0 & 0 & \ldots & \cos\theta & \ldots &  -\sin\theta & \ldots  & 0 \\
      & \ldots & \ldots & \ldots & \ldots &   \ldots & \ldots & \ldots & \ldots \\
    q & 0 & 0 & \ldots &  \sin\theta & \ldots & \cos\theta & \ldots  & 0 \\    
      & \ldots & \ldots & \ldots & \ldots &   \ldots & \ldots & \ldots & \ldots \\
      & 0 & 0 & \ldots & 0 & \ldots & 0  & \ldots  & 1 \\
  } \\
  {}
  \end{aligned}
\end{equation}
then the new estimator $\hat\beta^*$ is determined from the original estimator as follows
\begin{equation}
 \hat\beta_j^* = 
   \begin{cases}
      \hat\beta_p \cos\theta - \hat\beta_q \sin\theta & \mathit{for ~~} j = p \\
      \hat\beta_p \sin\theta + \hat\beta_q \cos\theta & \mathit{for ~~} j = q \\
      \hat\beta_j  & \mathit{otherwise}
  \end{cases}
\end{equation}
}

That is, if we rotate two components of the data points by an angle, the corresponding component of the estimator will be rotated by the same angle.

\subsection{Metamorphic relations} \label{section:MR}

\begin{figure*}[!ht]
	\centering
	\includegraphics[scale=0.25]{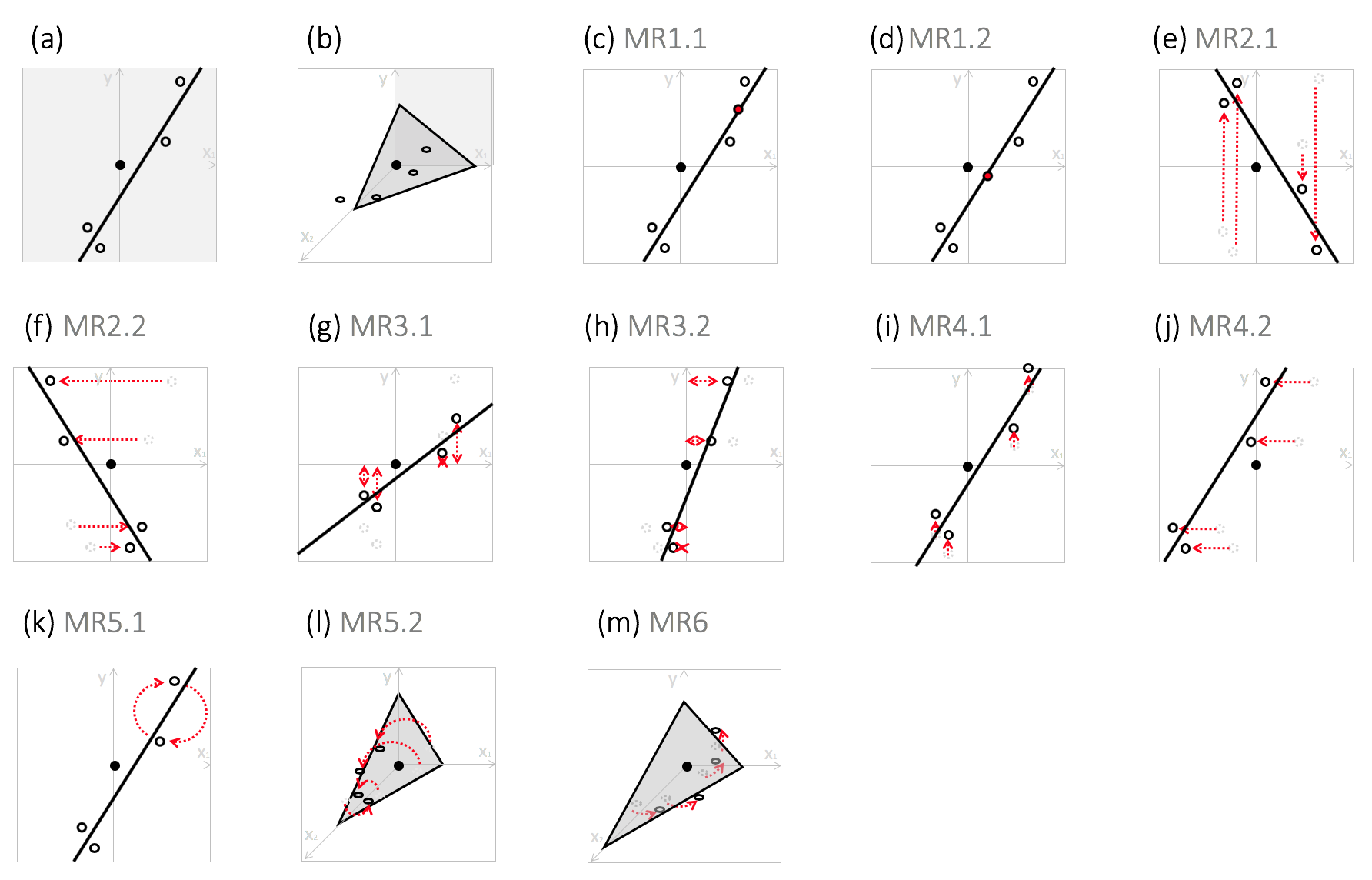} 
	\caption{Simplified original points and their regression line (hyperplane) in (a) two-dimensional and (b) three-dimensional views, and their transformations due to (c) MR1.1, (d) MR1.2, (e) MR2.1, (f) MR2.2, (g) MR3.1, (h) MR 3.2, (i) MR4.1, (j) MR4.2, (k) MR5.1, (l) MR5.2, and (m) MR6. The origin of coordinates is denoted by the black point. The dashed grey circles represent original points, i.e. the source inputs. The open black circles denote transformed points, which may overlay the original points if there is no transformation.  Newly added points are marked by the red circles. The follow-up inputs thus consist of transformed and added points. Black lines (grey hyperplanes) exhibit the follow-up regression lines (hyperplanes). The dashed red arrows sketch the transformation applied.}
	\label{figure:lines}
\end{figure*}

In this subsection, we will present 11 MRs for testing multiple linear regression systems. They are derived from properties presented in Section~\ref{section:Property} and are grouped into 6 categories. We will describe the regression form with intercept first, and discuss the regression form without intercept later. Table~\ref{table:mr_summary} summaries these MRs and their applicable types of regression, and Figure~\ref{figure:lines} illustrates the transformation examples for each MR.

\begin{table}[htbp]
\begin{adjustwidth}{-0.2cm}{}
\caption{Summary of MRs and their applicable types of regression}
\vspace{0.5cm}
\begin{center}
\scalebox{0.8}{
\begin{tabular}{llll}
\hline
\textbf{Category} 
	& \textbf{MR} & \textbf{Name}& \textbf{Form}$^{\mathrm{a}}$\\
\hline
    Unchanged & 1.1 &  
    		 Inserting a predicted point &
    		 I,C\\
    \hspace{0.2cm} predictions 
    & 1.2 &  
    		 Inserting the centroid &
    		 I\\
\hline
    Mirrored & 2.1 &  
    		 Reflecting the dependent variable &
    		 I,C\\
    \hspace{0.2cm} regression & 2.2 &  
    		 Reflecting an independent variable &
    		 I,C\\
\hline
    Scaled & 3.1 &  
    		 Scaling the dependent variable &
    		 I,C\\
    \hspace{0.2cm} regression & 3.2 &  
    		 Scaling an independent variable &
    		 I,C\\
\hline
    Shifted & 4.1 &  
    		 Shifting the dependent variable &
    		 I\\
    \hspace{0.2cm} regression & 4.2 &  
    		 Shifting an independent variable &
    		 I\\
\hline
    Reordered & 5.1 &  
    		 Swapping samples &
    		 I,C\\
    \hspace{0.2cm} regression & 5.2 &  
    		 Swapping two independent variables &
    		 I,C\\
\hline
    Rotated & 6 &  
    		 Rotating two independent variables &
    		 I,C\\
    \hspace{0.2cm} regression & &  
    		 &
    		 \\
\hline
\multicolumn{4}{l}{}\\
\multicolumn{4}{l}{
	$^{\mathrm{a}}$ 
	I: Regression with intercept as in Equation~(\ref{equation:intercept}) where ${\bf x}_0={\bf 1}_n$; }\\
\multicolumn{4}{l}{
	C: Constrained regression (without intercept) as in Equation~(\ref{equation:nointercept})}\\
\multicolumn{4}{l}{
	where ${\bf x}_0$ is an input variable whose data are provided by user. }\\
\end{tabular}
}
\label{table:mr_summary}
\end{center}
\end{adjustwidth}
\end{table}

Before we discuss the individual MR, let us introduce some notations for the regression with intercept. This regression is adopted as the default unless stated otherwise. First, let $I_s$ denote the source input of a linear regression system under test (SUT), and it consists of a set of $n$ data points
\begin{equation}
\begin{aligned}
	I_s &= \{ P_1^s, P_2^s, \ldots, P_n^s \} 
\end{aligned}
\end{equation}
where each data point $P_i^s $ ($i=1,2,\ldots,n$) can be denoted as
\begin{equation}
\begin{aligned}
	P_i^s =
		( x_{i,1}^s , x_{i,2}^s , \ldots , x_{i,d}^s , y_i^s )
\end{aligned}
\end{equation}
in which $x_{i,j}^s$ ($j=1,2,\ldots,d$) is the x-coordinate and $y_i^s$ is the y-coordinate of the point $P_i^s$. The output of the  SUT using the source input $I_s$ can be expressed by
\begin{equation}
\begin{aligned}
	O_s &= 
	\begin{bmatrix}
	\hat\beta^s_0 & \hat\beta^s_1 & \ldots & \hat\beta^s_d
	\end{bmatrix}^T
\end{aligned}
\end{equation}
The output $O_s$ is referred to as the source output. That is, the equation of the regression form with intercept is
\begin{equation}
	\hat{\bf y} =\hat\beta_0^s 
				+ {\bf x}_1 \hat\beta_1^s
				+ \ldots
				+ {\bf x}_d \hat\beta_d^s
\end{equation}
Let us denote the follow-up input $I_f$ be a set of $m$ data points, being generated using a given MR, that is 
\begin{equation}
\begin{aligned}
	I_f &= \{ P_1^f, P_2^f, \ldots, P_m^f \} 
\end{aligned}
\end{equation}
where 
\begin{equation}
\begin{aligned}
	P_i^f = 
		( x_{i,1}^f , x_{i,2}^f , \ldots , x_{i,d}^f , y_i^f )
\end{aligned}
\end{equation}
for $i=1,2,\ldots,m$. The output $O_f$ of the same program using the follow-up input $I_f$ is expressed by
\begin{equation}
\begin{aligned}
	O_f &= 
	\begin{bmatrix}
		\hat\beta^f_0 & \hat\beta^f_1 & \ldots & \hat\beta^f_d
	\end{bmatrix}^T
\end{aligned}
\end{equation}
The output $O_f$ is referred to as the follow-up output. For convenience in describing the MRs, we refer $I_s$ and $O_s$ ($I_f$ and $O_f$) to as the source (follow-up) data set and output, respectively.

In the context of MT, if $I_s$, $O_s$, $I_f$ and $O_f$ do not satisfy the relevant MR, the SUT is said to be faulty. Now, we are going to describe the MRs based on the previously mentioned properties, to be used in our experimentation to validate regression systems.

\subsubsection{Category 1: Unchanged predictions}
\vspace{0.25cm}

Two MRs in this category are based on Corollary~1, that is, adding a point predicted by the regression line into the original data set shall compute the same regression line. 

\vspace{0.25cm}
\noindent\textbf{MR1.1. }\textit{Inserting a predicted point}
\vspace{0.25cm}

\noindent The regression line will remain the same after being updated by adding a point selected arbitrarily from the line into the original data set. This MR is illustrated in Figure~\ref{figure:lines}c.

Given the source input $I_s$ and source output $O_s$, the follow-up input $I_f$ is formed by adding {a new} point $P^f_{n+1}$ to the source input $I_s$, that is
\begin{equation}
\begin{aligned}
	I_f &= I_s \cup \{ P^f_{n+1} \}
\end{aligned}
\end{equation}
where
\begin{equation}
\begin{aligned}
	P^f_{n+1} = 
		( x_{n+1,1}^f , x_{n+1,2}^f , \ldots , x_{n+1,d}^f , y_{n+1}^f )
\end{aligned}
\end{equation}
such that $x_{n+1,1}^f , x_{n+1,2}^f , \ldots , x_{n+1,d}^f$ are arbitrary real numbers, and $y^f_{n+1}$ is computed from these $x_{n+1,j}^f$ ($j=1,2,\ldots,d$) values and $O_s$ using the following equation
\begin{equation} \label{mr:inserting_predicted}
	y^f_{n+1} = \hat\beta^s_0
				+ x^f_{n+1,1} \hat\beta_1^s
				+ \ldots
				+ x^f_{n+1,d} \hat\beta_d^s 
\end{equation}
In other words, $P^f_{n+1}$ is a point that falls in the ``predicted'' regression equation. Then we expect to have the follow-up output be the same as the source output.
\begin{equation}
\begin{aligned} \label{mr:inserting_predicted_output}
	O_f &= O_s 	
\end{aligned}
\end{equation}

To sum up, the follow-up estimator is the same as the source estimator when a point predicted from the source model is included in the follow-up estimation. Instead of a single point, we can add several new points at the same time for the regression. The add-up will preserve the source estimators as long as the points are derived from the regression model. 

It is noted that, in practice, floating-point computations may give us slightly different values due to rounding. We thus need to consider the impact of round-off error in determining the violation of each MR. For example, instead of having the equality $\beta^s - \beta^f=0$ in this MR, the comparison should be relaxed by an inequality that accounts for an error tolerance. This will be explained in details in Subsection 4.3 (Assessment).

\vspace{0.25cm}
\noindent\textbf{MR1.2. }\textit{Inserting the centroid}
\vspace{0.25cm}

\noindent The linear regression line will remain the same after being updated by adding the centroid into the original data set for the intercept form, as illustrated in Figure~\ref{figure:lines}d. 
The centroid can be derived from the arithmetic mean of all values of its data points, which can be easily computed. This point belongs to the linear regression line. 

Given the source input $I_s$ and source output $O_s$, the follow-up input $I_f$ is formed by adding the centroid point $P^f_{n+1}$ to the source input $I_s$, that is
\begin{equation}
\begin{aligned}
	I_f &= I_s \cup \{ P^f_{n+1} \}
\end{aligned}
\end{equation}
where
\begin{equation}
\begin{aligned}
	P^f_{n+1} = 
		( x_{n+1,1}^f , x_{n+1,2}^f , \ldots , x_{n+1,d}^f , y_{n+1}^f )
\end{aligned}
\end{equation}
is such that $x_{n+1,1}^f , x_{n+1,2}^f , \ldots , x_{n+1,d}^f$ and $y_{n+1}^f$ are the averages of the corresponding values in the source data $I_s$, that is
\begin{equation}
\begin{aligned}
	y^f_{n+1} &= \frac{1}{n} (y_1^s + y_2^s + \ldots + y_n^s) \\
	x^f_{n+1,j} &= \frac{1}{n} ( x_{1,j}^s + x_{2,j}^s + \ldots + x_{n,j}^s ) 
\end{aligned}
\end{equation}
for $j=1,2,\ldots,d$. If the regression program is configured to run with intercept, then we have
\begin{equation}
\begin{aligned}
	O_f &= O_s 	
\end{aligned}
\end{equation}

In other words, adding the centroid of data into the source input to form a follow-up input will not change the follow-up estimator for the regression form with intercept. 



\subsubsection{Category 2: Mirrored regression}
\vspace{0.25cm}

The MRs in this category are based on Corollary~3 and Corollary~4 with the scaling parameter be set as $-1$. This set of MRs is related to how reflecting the data points will reflect the regression line accordingly. 

\vspace{0.25cm}
\noindent\textbf{MR2.1. }\textit{Reflecting the dependent variable}
\vspace{0.25cm}

\noindent Reflecting the points over a certain $x$-axis will reflect the regression line over the same axis, as illustrated in Figure~\ref{figure:lines}e.

Given the source input $I_s$ and source output $O_s$, the follow-up input $I_f$ consists of $n$ points 
\begin{equation}
\begin{aligned}
	I_f &= \{ P_1^f, P_2^f, \ldots, P_n^f \} 
\end{aligned}
\end{equation}
where for each $P_i^f$ ($i=1,2,\ldots,n$), the value of its x-coordinate remains unchanged (that is $x_{i,j}^f = x_{i,j}^s$, $j=1,2,\ldots,d$) and its y-coordinate is reflected, that is
\begin{align}
  y^f_i = - y_i^s
\end{align}
Then, we have
\begin{equation}
\begin{aligned}
	O_f &= -O_s
\end{aligned}
\end{equation}

That is, the follow-up estimator is a reflection of the source estimator when the sign of the dependent variable is reversed in the follow-up input set.

\vspace{0.25cm}
\noindent\textbf{MR2.2. }\textit{Reflecting an independent variable while keeping the others unchanged}
\vspace{0.25cm}

\noindent Reflecting the points over the y-axis will reflect the regression line over the same axis, as illustrated in Figure~\ref{figure:lines}f.

Given the source input $I_s$ and source output $O_s$, the follow-up input $I_f$ is defined to consist of $n$ points 
\begin{equation}
\begin{aligned}
	I_f &= \{ P_1^f, P_2^f, \ldots, P_n^f \} 
\end{aligned}
\end{equation}
where for each $P_i^f$ ($i=1,2,\ldots,n$), the x-coordinate of an independent variable, say ${\bf x}_k$, is reflected while the x-coordinates of other independent variables and the y-coordinates remain the same, that is
\begin{equation}
 x^f_{i,j} =
     \begin{cases}
 		 -x_{i,k}^s &\quad\text{for } j = k \\
		 x_{i,j}^s &\quad\text{for } j=1,2,\ldots,d \text{ and } j \neq k
      \end{cases}
\end{equation}
The follow-up output $O_f$ 
\begin{equation}
\begin{aligned}
	O_f &= 
	\begin{bmatrix}
		\hat\beta^f_0 & \hat\beta^f_1 & \ldots & \hat\beta^f_d
	\end{bmatrix}^T
\end{aligned}
\end{equation}
is determined by the following equation 
\begin{equation}
  \hat\beta^{f}_{k} = 
     \begin{cases}
       -\hat\beta_k^s &\quad\text{for } j = k \\
       \hat\beta_j^s &\quad\text{for } j=0,1,\ldots,d \text{ and } j \neq k \\
     \end{cases}
\end{equation}

In a nutshell, if the sign of an independent variable is reversed, the sign of corresponding component of the estimator should also be reversed.

\vspace{0.25cm}
\subsubsection{Category 3: Scaled regression}
\vspace{0.25cm}

The MRs in this category are also based on Corollary~5 (MR3.1) and Corollary~6 (MR3.2) for an arbitrary positive scaling factor. When we scale particular coordinates of data points along an axis by a positive factor, the slope of regression line will be scaled accordingly. 

\vspace{0.25cm}
\noindent\textbf{MR3.1. }\textit{Scaling the dependent variable}
\vspace{0.25cm}

\noindent After the points are scaled in the y axis by a given factor, the regression line will be scaled by the same ratio, as shown in Figure~\ref{figure:lines}g.

Given the source input $I_s$ and source output $O_s$, the follow-up input $I_f$ is defined to consist of $n$ points 
\begin{equation}
\begin{aligned}
	I_f &= \{ P_1^f, P_2^f, \ldots, P_n^f \} 
\end{aligned}
\end{equation}
where for each $P_i^f$ ($i=1,2,\ldots,n$), the values of its x-coordinates remain unchanged ($x_{i,j}^f = x_{i,j}^s$, $j=1,2,\ldots,d$) and its y-coordinate is scaled by a factor $a$ ($a>0$), that is
\begin{align}\label{mr:scale_y}
  y^f_i = a~y_i^s
\end{align}
Then, the sign of follow-up output $O_f$ would be scaled by the same {factor}
\begin{equation}
\begin{aligned}\label{mr:scale_y_result}
	O_f &= a~O_s
\end{aligned}
\end{equation}

\noindent In other words, when values of the dependent variable is scaled by a given factor, all components of the estimator will be scaled proportionally by the same factor.

\vspace{0.25cm}
\noindent\textbf{MR3.2. }\textit{Scaling an independent variable while keeping the others unchanged}
\vspace{0.25cm}

\noindent After the points are scaled in a particular x axis by a given factor, the slope of regression line will be scaled reciprocally with respect to that axis, as illustrated in Figure~\ref{figure:lines}g.

Given the source input $I_s$ and source output $O_s$, the follow-up input $I_f$ is defined to consist of $n$ points 
\begin{equation}
\begin{aligned}
	I_f &= \{ P_1^f, P_2^f, \ldots, P_n^f \} 
\end{aligned}
\end{equation}
where for each $P_i^f$ ($i=1,2,\ldots,n$), the x-values of an independent variable, say ${\bf x}_k$, are scaled by a factor $b$ ($b>0$), while the values of other independent variables and the dependent variable remain the same, that is
\begin{equation}
 x^f_{i,j} =
     \begin{cases}
 		 b~x_{i,k}^s &\quad\text{for } j = k \\
		 x_{i,j}^s &\quad\text{for } j=1,2,\ldots,d \text{ and } j \neq k
      \end{cases}
\end{equation}
and
\begin{equation}
   y^f_i  = y_i^s
\end{equation}
The follow-up output $O_f$
\begin{equation}
\begin{aligned}
	O_f &= 
	\begin{bmatrix}
		\hat\beta^f_0 & \hat\beta^f_1 & \ldots & \hat\beta^f_d
	\end{bmatrix}^T
\end{aligned}
\end{equation}
would be defined by the following equation 
\begin{equation}
  \hat\beta^{f}_{k} = 
     \begin{cases}
       \frac{1}{b}\hat\beta_k^s &\quad\text{for } j = k \\
       \hat\beta_j^s &\quad\text{for } j=0,1,\ldots,d \text{ and } j \neq k \\
     \end{cases}
\end{equation}

In other words, if the values of an independent variable are scaled by a given factor, the corresponding component of the predicted estimator would be scaled by the reciprocal of the scaling constant.

\vspace{0.25cm}
\subsubsection{Category 4: Shifted regression}
\vspace{0.25cm}

The MRs in this category are based on Corollary~7 and Corollary~8, and applicable only for the regression with intercept. The intercept of the regression line will accumulate all modified distances if you shift the points along an axis.

\vspace{0.25cm}
\noindent\textbf{MR4.1. }\textit{Shifting the dependent variable}
\vspace{0.25cm}

\noindent When the points are shifted by a given distance along the y axis, the regression line will be shifted along this axis by the same distance, as illustrated in Figure~\ref{figure:lines}i.

Given the source input $I_s$ and source output $O_s$, the follow-up input $I_f$ {is defined to consist} of $n$ points 
\begin{equation}
\begin{aligned}
	I_f &= \{ P_1^f, P_2^f, \ldots, P_n^f \} 
\end{aligned}
\end{equation}
where for each $P_i^f$ ($i=1,2,\ldots,n$), the value of its y-coordinate is shifted by a distance $a$
\begin{equation}
 y^f_{i} = y_{i}^s + a 
\end{equation}
{Then,} the follow-up output $O_f$ 
\begin{equation}
\begin{aligned}
	O_f &= 
	\begin{bmatrix}
		\hat\beta^f_0 & \hat\beta^f_1 & \ldots & \hat\beta^f_d
	\end{bmatrix}^T
\end{aligned}
\end{equation}
would be defined by the following equation 
\begin{equation}
 \hat\beta^{f}_j = 
     \begin{cases}
       \hat\beta_0^s + a &\quad\text{for } j = 0 \\
       \hat\beta_j^s &\quad\text{for } j = 1,2,\ldots,d \\
     \end{cases}
\end{equation}

In brief, if a constant is added into the values of dependent variable, the intercept of the new estimator would be increased by the same value.

\vspace{0.25cm}
\noindent\textbf{MR4.2. }\textit{Shifting an independent variable while keeping the others unchanged}
\vspace{0.25cm}

\noindent When the points are shifted by a given distance along a certain x axis, the regression line will be shifted in parallel along this axis accordingly, as illustrated in Figure~\ref{figure:lines}j.

Given the source input $I_s$ and source output $O_s$, the follow-up input $I_f$ is defined to consist of $n$ points 
\begin{equation}
\begin{aligned}
	I_f &= \{ P_1^f, P_2^f, \ldots, P_n^f \} 
\end{aligned}
\end{equation}
where for each $P_i^f$ ($i=1,2,\ldots,n$), the value of an independent variable, say ${\bf x}_k$, is shifted by a distance $b$ while the x-coordinates of other independent variables and the y-coordinates remain the same, that is
\begin{equation}
 x^f_{i,j} =
     \begin{cases}
 		 x_{i,k}^s + b &\quad\text{for } j = k \\
		 x_{i,j}^s &\quad\text{for } j=1,2,\ldots,d \text{ and } j \neq k
      \end{cases}
\end{equation}
The follow-up output $O_f$
\begin{equation}
\begin{aligned}
	O_f &= 
	\begin{bmatrix}
		\hat\beta^f_0 & \hat\beta^f_1 & \ldots & \hat\beta^f_d
	\end{bmatrix}^T
\end{aligned}
\end{equation}
would be defined by the following equation 
\begin{equation}
  \hat\beta^{f}_{j} = 
     \begin{cases}
       \hat\beta_0^s - b \hat\beta_k^s &\quad\text{for } j = 0 \\
       \hat\beta_j^s &\quad\text{for } j = 1,2,\ldots,d \\
     \end{cases}
\end{equation}

So, if a constant is added into values of an independent variable, the intercept component of follow-up estimator would be decreased by an amount equal to the product of the constant and the value of the corresponding component of source estimator.

\vspace{0.25cm}
\subsubsection{Category 5: Reordered regression}
\vspace{0.25cm}

\noindent The MRs in this category are associated with Corollary~9 and Corollary~10, such that reordering the axes of data points may need changes of the order of axes of the regression hyperplane.

\vspace{0.25cm}
\noindent\textbf{MR5.1. }\textit{Swapping samples}
\vspace{0.25cm}

\noindent Swapping any two data points does not alter the regression hyperplane, as illustrated in Figure~\ref{figure:lines}k.

Given the source input $I_s$ and source output $O_s$, suppose that we swap two data points, say $P_p$ and $P_q$ ($1\leq p, q\leq d$), to define the follow-up input $I_f$ which consists of $n$ points 
\begin{equation}
\begin{aligned}
	I_f &= \{ P_1^f, P_2^f, \ldots, P_n^f \} 
\end{aligned}
\end{equation}
where
\begin{equation}
 P_i^f = 
     \begin{cases}
        P_{p}^s &\quad\text{for } i = q\\     
        P_{q}^s &\quad\text{for } i = p\\     
        P_{i}^s &\quad\text{otherwise}
     \end{cases}
\end{equation}
Then, the follow-up output $O_f$ would be the same as the source input $O_s$, that is
\begin{equation}
\begin{aligned}
	O_f &= O_s
\end{aligned}
\end{equation}

As a result, we would have the same estimator no matter how we change the order of data points for the regression. 

\vspace{0.25cm}
\noindent\textbf{MR5.2. }\textit{Swapping two independent variables while keeping the others unchanged}
\vspace{0.25cm}

\noindent Swapping two axes does not actually change the essence of regression hyperplane in individual axes, as illustrated in Figure~\ref{figure:lines}l.

Assume that two axes of the data points, say ${\bf x}_p$ and ${\bf x}_q$ ($1\leq p, q\leq d$) are swapped. Given the source input $I_s$ and source output $O_s$, the follow-up input $I_f$ is defined to consist of $n$ points 
\begin{equation}
\begin{aligned}
	I_f &= \{ P_1^f, P_2^f, \ldots, P_n^f \} 
\end{aligned}
\end{equation}
where for each $P_i^f$ ($i=1,2,\ldots,n$), the values of its coordinates are determined as follows
\begin{equation}
 x^f_{i,j} = 
     \begin{cases}
        x_{i,p}^s &\quad\text{for } j = q\\     
        x_{i,q}^s &\quad\text{for } j = p\\     
        x_{i,j}^s &\quad\text{otherwise}
     \end{cases}
\end{equation}
and
\begin{equation}
   y^f_i  = y_i^s
\end{equation}
Then, the follow-up output $O_f$
\begin{equation}
\begin{aligned}
	O_f &= 
	\begin{bmatrix}
		\hat\beta^f_0 & \hat\beta^f_1 & \ldots & \hat\beta^f_d
	\end{bmatrix}^T
\end{aligned}
\end{equation}
would be defined as follows
\begin{equation}
 \hat\beta^{f}_{j} = 
     \begin{cases}
        \hat\beta_{p}^s &\quad\text{for } j = q\\     
        \hat\beta_{q}^s &\quad\text{for } j = p\\     
        \hat\beta_{j}^s &\quad\text{otherwise}
     \end{cases}
\end{equation}

In other words, swapping two independent variables while keeping the others unchanged will only swap the two relevant components of the follow-up estimator.

\vspace{0.25cm}
\subsubsection{Category 6: Rotated regression}
\vspace{0.25cm}

This MR is based on Corollary~11, which specifies how the regression hyperplane will change after rotating axes related to independent variables.

\vspace{0.25cm}
\noindent\textbf{MR6. }\textit{Rotating two independent variables while keeping the others unchanged}
\vspace{0.25cm}

\noindent When the points are rotated in the plane perpendicular to the y axis, the regression hyperplane will be rotated accordingly, as illustrated in Figure~\ref{figure:lines}k.

Suppose that the axes of two independent variables, say ${\bf x}_p$ and ${\bf x}_q$ ($1\leq p, q\leq d, p \neq q$), are rotated by an angle $\theta$ in the counter-clockwise direction. Given the source input $I_s$ and source output $O_s$, the follow-up input $I_f$ is defined to consist of $n$ points 
\begin{equation}
\begin{aligned}
	I_f &= \{ P_1^f, P_2^f, \ldots, P_n^f \} 
\end{aligned}
\end{equation}
where for each $P_i^f$ ($i=1,2,\ldots,n$), the values of its x-coordinates are defined by the following equation
\begin{equation}
 x^f_{i,j} = 
     \begin{cases}
        x_{i,p}^s \cos\theta - x_{i,q}^s \sin\theta &\quad\text{for } j = p\\     
        x_{i,p}^s \sin\theta + x_{i,q}^s \cos\theta &\quad\text{for } j = q\\     
        x_{i,j}^s &\quad\text{otherwise}
     \end{cases}
\end{equation}
and the values of its y-coordinate remain the same, that is
\begin{equation}
   y^f_{i} = y^s_{i} 
\end{equation}
Then, the follow-up output $O_f$
\begin{equation}
\begin{aligned}
	O_f &= 
	\begin{bmatrix}
		\hat\beta^f_0 & \hat\beta^f_1 & \ldots & \hat\beta^f_d
	\end{bmatrix}^T
\end{aligned}
\end{equation}
would be defined by the following equation 
\begin{equation}
  \hat\beta^{f}_{j} = 
     \begin{cases}
        \hat\beta_{p}^s \cos\theta - \hat\beta_{q}^s \sin\theta &\quad\text{for } j = p\\     
        \hat\beta_{p}^s \sin\theta + \hat\beta_{q}^s \cos\theta &\quad\text{for } j = q\\     
        \hat\beta_{j}^s &\quad\text{otherwise}
     \end{cases}
\end{equation}

Basically, if we rotate the axes of any two independent variables by an angle, the corresponding components of estimator would be also ``rotated'' by the same angle.

\subsection{Testing scheme}

The process to test the linear regression system using MT is shown in Figure~\ref{figure:testingflow}. The steps are summarized as follows:

\begin{figure*}[!ht]
	\centering
	\includegraphics[scale=0.42]{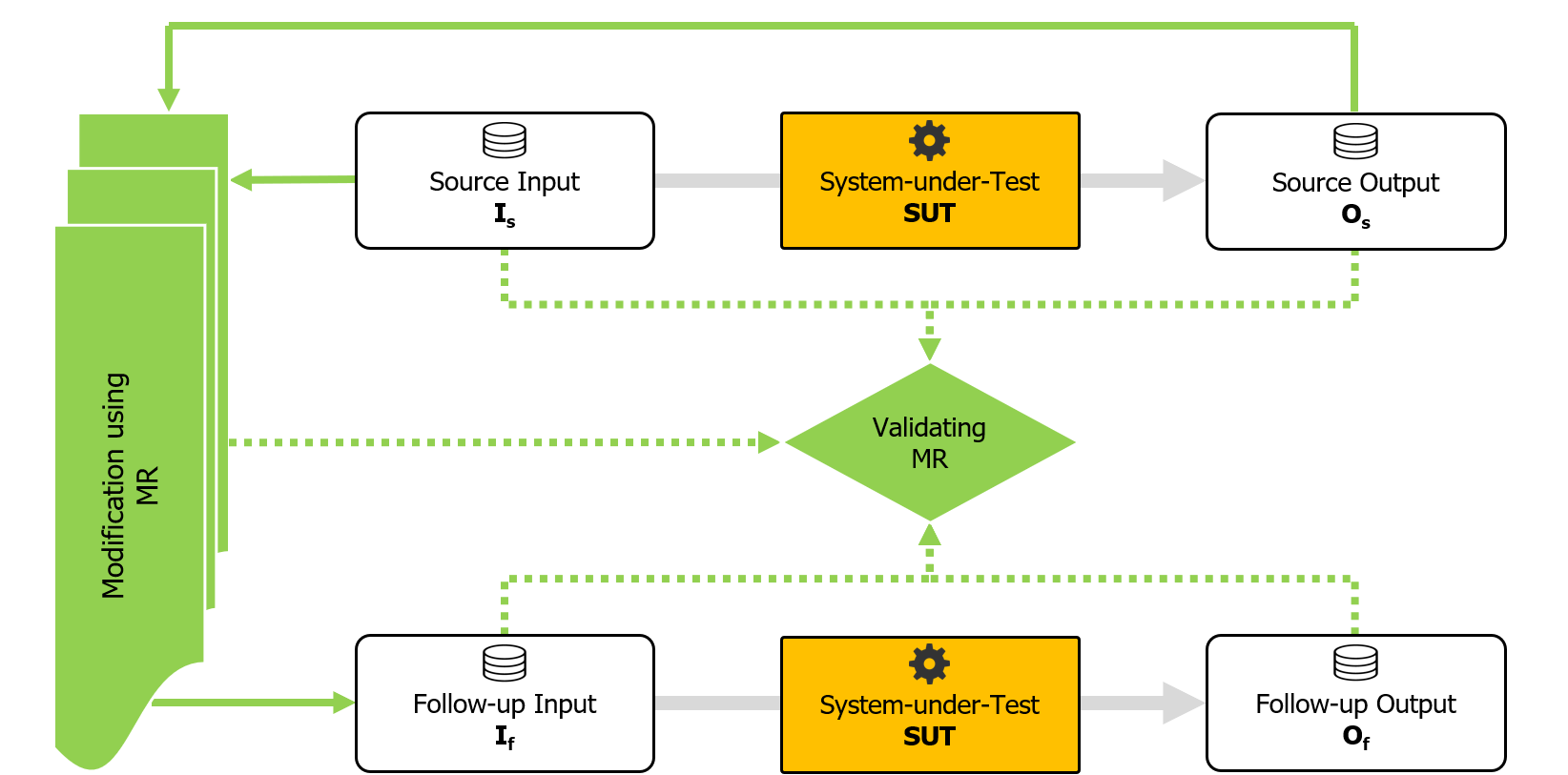} 
	\caption{The process of testing program using MT. After the first execution of SUT, source input and source output are used to generate the follow-up input with reference to the given MR. All inputs and outputs are then used to validate MR.}
	\label{figure:testingflow}
\end{figure*} 

\begin{enumerate}
\item \textit{Preparing source input}: Preparing the source input $I_s$ to test the SUT. In testing the regression system, the source input consists of data points associated with independent and dependent variables ${x_1^s, x_2^s,\ldots,x_d^s,y^s}$.
\item \textit{Obtaining source output}: After executing the SUT using the source input, we obtain the source output $O_s$. Here, the source output contains $(\beta_0^s,\beta_1^s,\ldots,\beta_d^s)$. 
\item \textit{Generating follow-up input}: We generate the follow-up input $I_f$ from the source-input $I_s$ and the source-output $O_s$ with the reference to the given MR. For instance, we can create a follow-up input by adding a new point, say $(1,4)$ predicted from the regression line $y=x+3$ as given in the forth paragraph in Section~1 as an example. Now that the follow-up input has all these 4 points: $(2,4), (7,6), (9,10)$ and $(1,4)$.
\item \textit{Obtaining follow-up output}: We execute the SUT using the follow-up input to get the follow-up output $O_f$. For the regression program, the follow-up output is $(\beta_0^f,\beta_1^f,\ldots,\beta_d^f)$. 
\item \textit{Validating MR}: We compare $I_s, O_s, I_f$ and $O_f$ against the MR. If it is violated, the SUT is concluded to be faulty. As per above example, we require the follow-up output $O_f$ and source output $O_s$ to have $\hat\beta^f_i=\hat\beta^s_i$ for $i={0,1,\ldots,d}$ according to Equation (\ref{mr:inserting_predicted_output}) for MR1.1 while in practice we may further consider the computational error for asserting this equality. In general, all sets $I_s, O_s, I_f$ and $O_f$ can be adopted to validate the MR.
\end{enumerate}

\section{Experiments}

We examine the applicability of our MRs to test Scikit-learn, a well-known open-source library. This library is currently adopted by 112000 registered users in GitHub. Its latest stable version is 0.23, which consists of the multiple linear regression, named LinearRegression() class. The library consists of a fault about its mishandling of dataframe that has column labels to denote variable names and rows to store data. It happened when users train a model with a dataframe and adopt the model for prediction or retraining using a dataframe having the same column names but in a different order. This fault affects linear regression, as well as other linear models in Scikit-learn.  From the user's perspective, they may be unaware of this issue since most of them do not have a deep understanding about the regression program. Some users were only aware of the problem until they examined the code after realising that their predictions are contrary to their understandings. This defect has been reported since 2016 and has not been fixed yet \citep{github1,github2,github3}.

We believe that our MR5.2 is able to reveal this problematic issue. We actually have performed a simple experiment to confirm this. We first prepare a dataframe of trained data as our source test case. We then use this source dataframe in the relevant modules in Scikit-learn for training and adopt the trained model for prediction. Follow-up test cases are then prepared by swapping two columns in the original source dataframe. As this is a follow-up dataframe for prediction, the column for the ``predicted value'' is omitted from the dataframe whereas the columns for the ``independent variables'' stay. After adopting the follow-up dataframe for prediction, we found that the predicted values from the Scikit-learn are inconsistent because the source dataframe  used for modelling and the follow-up dataframe used for prediction have different orderings in their columns. We would like to argue that this problem will be revealed if the developers apply our MRs (in particular MR5.2) to test their code during development.

In other words, the MR developed for testing the regression algorithm is able to detect this problem of Scikit-learn even though the blunder is not related to its regression calculation.

\subsection{System under test and mutation analysis} 

To further measure the effectiveness of these identified MRs, mutation analysis is applied in this study. Since our focus is to test the implementation of linear regression algorithm, mutation analysis is a good option because it allows us to evaluate the effectiveness of MT against a wide range of possible faults. It should be noted that we have not customized the testing to a particular system. As explained earlier in the introduction, while testing such system is important, it requires a particular set of features and properties specific to the system that may not be applicable to be used to test other ones. Moreover, {each real-life system} may subject to a specific type of faults, which is unable to give us a comprehensive overview of effectiveness of MT.

For the mutation analysis, we used {five} {C/C++} regression programs, referred to as {\it Press}, {\it Vijayan}, {\it Oscar}, {\it Quinn-Curtis}, and {\it Barr}. {\it Press} is a standard regression program provided in the well-known C++ textbook ``Numerical Recipes 3rd Edition: The Art of Scientific Computing'' \citep{press2007}. Published by William H. Press and colleagues, the source program consists of \textit{fitsvd.h} and \textit{svd.h}, having a total 430 lines of code (LOC). Two programs, namely {\it Vijayan} and  {\it Oscar}, are from the GitHub repository. {\it Vijayan} is {a C} program developed by Vijayan Thanusan \citep{vijayan}, having a total number of 350 LOC.  {\it Oscar} is coded by Oscar Hamilton, consisting of 3 different files \textit{main.c}, \textit{matrix.c} and \textit{matrix.h}, and having a total number of 650 LOC \citep{oscar}. Another two programs are from the websites of two academic institutions, {\it Quinn-Curtis} from Pazmany Peter Catholic University (Hungary) and Southern Methodist University (United States). {\it Quinn-Curtis} has 430 LOC and originally supplied by Quinn-Curtis company \citep{quinn-curtis}; while {\it Barr} program has 590 LOC made by Richard S. Barr \citep{barr}. All programs are then complied using the gcc 4.2.1 associated with with Apple LVM version 10.0.0 (clang-1000.11.45.5), and run on a system of iMac computer (OS version 10.13.6).

Given a program, if we modify the original version by making a small syntactic change, we obtain a mutant of the original program. The small syntactic change is usually referred to as a mutation operator. Table~\ref{table:mutants} lists the mutation operators we used to generate the mutants of our subject program. These mutation operators mimic typical programming mistakes. We developed a tool to systematically apply the mutation operators in Table~\ref{table:mutants}, one at a time, to generate the mutants for our experiments. Each mutant is a result of one application of the mutation operators.

\begin{table*}[tbp] 
\begin{adjustwidth}{-0.4cm}{}
\caption{Summary of generated mutants, compilable mutants and non-equivalent mutants used for testing, being classified by mutated keywords.}
\vspace{0.5cm}
\begin{center}
{
\scalebox{0.8}{
\begin{tabular}{|l|l|l|r|r|r|}
\hline
\textbf{Mutation category} 
	& \textbf{Mutated keywords}
	& \textbf{Description}
	& \textbf{Generated} 
	& \textbf{Compilable}
	& \textbf{Used}\\
\hline
    array\_construct &
       ( ), (*0), (*(-1)), (*2)  & Mistaken construction of array & 498    &     293    &     98   \\ 
    array\_index &
       [], [-1+], [1+], [0*]   & Misplaced array index & 1512    &     1506    &     894   \\ 
    array\_swap1 $^{\mathrm{a}}$ &
        [i], [j], [k], [0], [1]  &  Misplaced element of one-dimensional array & 1452    &     1222    &     710   \\ 
    array\_swap2 $^{\mathrm{a}}$ &
        [i][j], [j][i], [i][k], [k][i], [j][k],[k][j]   
        	& Misplaced elements of multi-dimensional array & 505    &     412    &     292   \\ 
    condition\_if &
        if(), if (!), if(true$\vert\vert$), if(false\&\&)  
        	&  Illogical branching condition & 45    &     24    &     16   \\ 
    condition\_index &
        i=, j=, k=   &  Misused counter in loops & 200    &     170    &     111   \\ 
    condition\_loop &
        break, continue, \{;\}   &  Discontinuation of loops & 14    &     14    &     7   \\ 
    data\_complex $^{\mathrm{b}}$ &
        Doub, MathDoub/I, VecDoub/I  & Wrong data type for numeric array & 472    &     6    &     0   \\ 
    data\_simple $^{\mathrm{b}}$ &
        Doub, Int  &  Wrong data type for numeric variable &  94    &     23    &     11   \\ 
    function\_parameter &
       $f$(), $f$(*0), $f$(*(-1)), $f$(*2)  & Mistaken use of function numeric parameter & 1832    &     824    &     356   \\ 
    function\_return &
        return 0, 1, -1, 2, -1*, NULL  & Wrong value of function's return & 168    &     103    &     39   \\ 
    logic\_combination &
       	\text{$\vert$, \&\&, \&, $\vert\vert$, \&\& !}  
       		& Illogical combination of conditions & 430    &     90    &     45   \\ 
    logic\_comparison &
        $!=,<,>,<=,>=,==$  & Mistaken logical operators & 1575    &     1336    &     564   \\ 
    logic\_disable &
        ?, $\&\&$ false ? $,\vert\vert$ true ?   & Mistaken logical return & 14    &     14    &     1   \\ 
    logic\_not &
        !, \~  & Mistaken negation operator & 31    &     8    &     5   \\ 
    logic\_value &
        true, false   & Mistaken logical condition & 2    &     2    &     0   \\ 
    math\_increment &
        ++, -~-, +=2, -=2  & Misused incremental operator & 600    &     522    &     223   \\ 
    math\_initial &
        =0, =1, =2   & Wrong assignment of initial value & 226    &     226    &     96   \\ 
    math\_operator $^{\mathrm{c}}$ &
        +, -, *, /, \%   & Wrong use of mathematical operators &  4800    &     2097    &     840   \\ 
    math\_values $^{\mathrm{c}}$ &
        +1, -1, +2, -2   &  Mistaken incremental values &  285    &     285    &     122   \\ 
\hline
    \multicolumn{3}{|r|}{\textbf{Total}}
	& 14755
	& 9177
	& 4430 \\ 
\hline	
\multicolumn{5}{l}{$^{\mathrm{a,c}}$When a mutant belongs to two groups at the same time, it will be manually categorized. A mutant belonged to both  
array\_swap1}\\
\multicolumn{5}{l}{and array\_swap2 will be only classified as array\_swap2; math\_operator and math\_values will be only classified as math\_values.}\\
\multicolumn{5}{l}{$^{\mathrm{b}}$ Customized data types for matrix, vector, floating and integer numbers \citep{press2007}.}
\end{tabular}
} 
}
\label{table:mutants}
\end{center}
\end{adjustwidth}
\end{table*}

After using the mutation generation tool, we have obtained a total number of 14755 mutants from 5 different original programs, namely {\it Press}, {\it Vijayan}, {\it Oscar}, {\it Quinn-Curtis}, and {\it Barr} (Table~\ref{table:put_summary}). Among these mutants, only 9177 of them can be successfully compiled using the system's {gcc} 4.2.1 compiler (Table~\ref{table:put_summary}). Their categories of mutation types are presented in (Table~\ref{table:mutants}).

\begin{table}[htbp]
\begin{adjustwidth}{+0.2cm}{}
\caption{Summary of programs and number of mutants generated, compilable and non-equivalent (used)}
\vspace{0.5cm}
\begin{center}
\scalebox{0.8}{
\begin{tabular}{l r r r}
\hline
\textbf{Program} 
	& \textbf{Generated} & \textbf{Compilable}
	& \textbf{Non-equivalent}\\
\hline
 Press   & 6127   & 3592   & 2299   \\ 
 Vijayan   & 2163   & 1226   & 439   \\ 
 Oscar   & 1419   & 1213   & 686   \\ 
 Quinn-Curtis   & 2803   & 1533   & 446   \\ 
 Barr   & 2243   & 1613   & 560   \\ 
\hline
 Total   & 14755   & 9177   & 4430   \\ 
\hline
\end{tabular}
}
\end{center}
\label{table:put_summary}
\end{adjustwidth}
\end{table}

Another major issue of mutation analysis is to eliminate the equivalent mutants, or more precisely, the mutants that are equivalent to the {original programs}. A mutant is equivalent to the original program when the outputs of the mutant and the original program are exactly the same for all possible inputs. There are usually two methods to determine a mutant is equivalent to the original program. The first one is manual inspection. For this method, manually inspecting 9177 mutants, one at a time, to judge whether the inspected mutant is equivalent to the original program is too time consuming and resource intensive. The second method is to compare whether the outputs of the mutant and the original program are the same for all possible input values. As for linear regression program, there are infinitely many inputs. Hence, this second method is still practically infeasible. Nonetheless, we adopt a working definition of an equivalent mutant: {\textit A mutant is equivalent to its original (linear regression) program if the outputs of the mutant and the original program are the same for 100 randomly generated inputs.}

\begin{figure}[ht]
	\centering
	\includegraphics[scale=0.58]{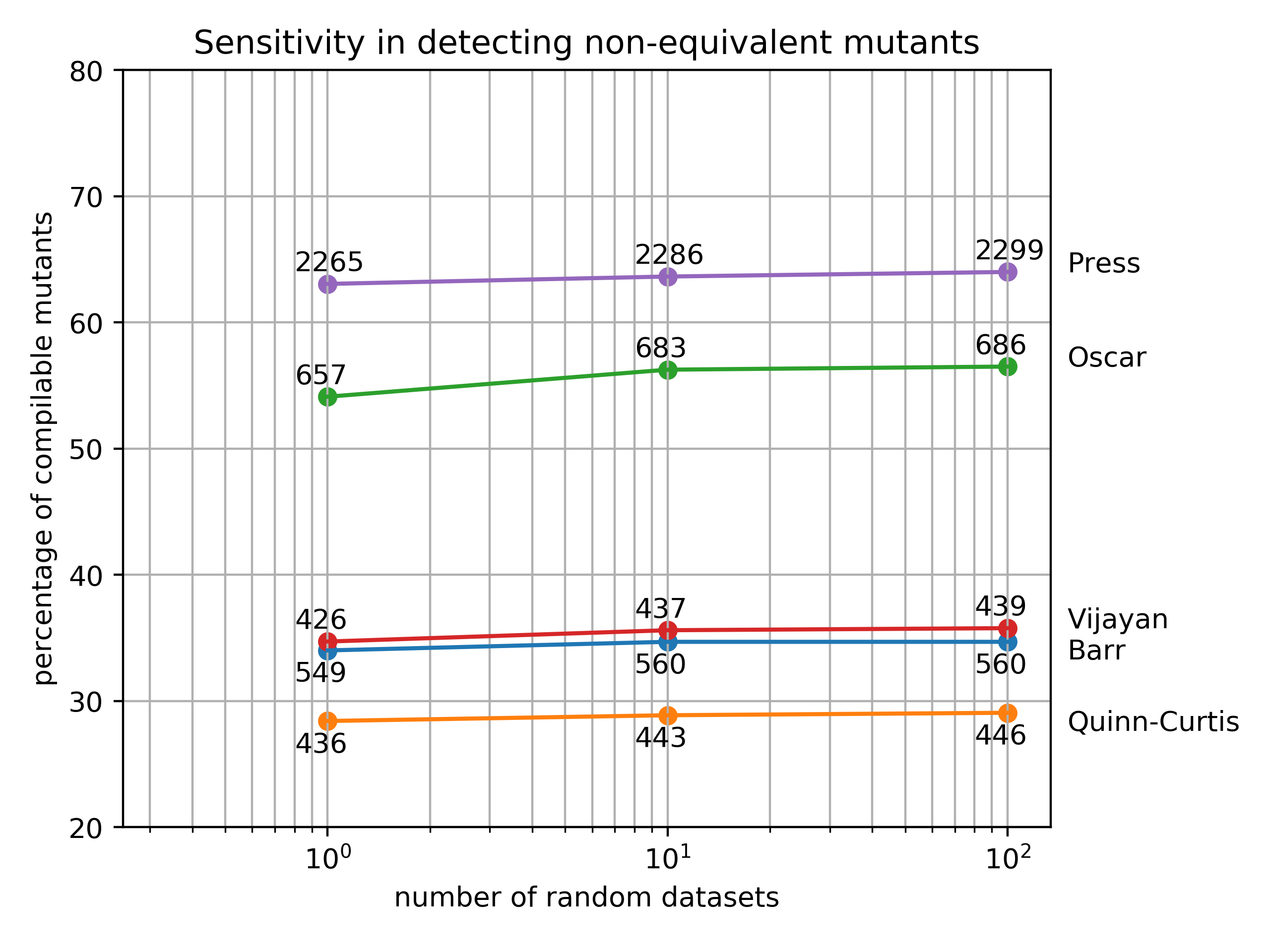}
	\caption{{Detection of non-equivalent mutants from different programs.}}
	\label{figure:equivalent}
\end{figure}

Our working definition uses ``100 randomly generated inputs'' to determine the equivalence rather than using ``all possible inputs''. We chose 100 because we have performed a sensitivity analysis on the effect of the number of remaining non-equivalent mutants versus the number of randomly generated inputs. Figure~\ref{figure:equivalent} plots the number of remaining non-equivalent mutants versus the number of randomly generated inputs from 1 to 100. Please be reminded that, for linear regression program, the input is a set of data points for linear regression calculations. We can see that the number of remaining non-equivalent mutants changes from 2265 (63.1\% of 3592) with 1 randomly generated input to 2286 (63.6\%) with 10, to 2299 (64.0\%) with 100 for the {\it Press} program (Figure~~\ref{figure:equivalent}). In general, we observed from Figure~$\ref{figure:equivalent}$ that there are 21 more non-equivalent mutants if we adopt 100 randomly generated inputs instead of 10 for all programs. This number accounts for only 0.2\% of the total number of compilable mutants. We anticipated that there will still be some non-equivalent mutants found if we use 1000 instead of 100 randomly generated inputs in our definition. However, the additional detections will be very minimal and negligible based on the trends that we observed in Figure~$\ref{figure:equivalent}$. Using our working definition of "equivalent" mutant, we can further remove 4747 mutants from the successfully compiled 9177 mutants. As a result, we have 4430 non-equivalent mutants for our experiments.

\subsection{Datasets}

We have generated 100 source datasets for the mutation analysis. The size of a dataset used for testing is basically characterized by two factors: the number of independent variable ($d$) and the number of sampling points ($n$). We randomly choose the numbers of independent variables $d$ in between 2 and 16, and the number of samples in between 20 and 200. These values are adopted not only because they are often used in practice \citep{wu2017, luu2015, luu2018}, but also because of the availability of resources.

Then the independent variables $x_1, x_2,\ldots, x_d$ and the dependent $y$ are randomly generated using normal distributions. Each one of these variables, say $z$, is a set of $n$ points $z_i$ ($i=1,2,\ldots,n$) that have the mean value $\overline{z}$ and the deviations $z_i'$
\begin{align}\label{equation:variant}
	z_i &= \overline{z} + z_i' \\
	z_i' &= Z \times r \times nrand
\end{align}
where $nrand()$ is a function to generate random numbers in a normal distribution with mean 0 and variance 1. The range $Z$ is a fixed value for each $z$, being selected randomly in between from 0 to 100. The maximum signal-to-noise ratio ($r$) is set as $1/10$ (that is,  10\%). Such a configuration allows some independent variables to vary significantly, while keeping others more stable. The data $x_{ij}$ for the variable $x_j$ ($j=1,2,\ldots,d$) are generated by 
\begin{align}
	x_{ij} =  \overline{x_j} + x_{ij}' 
\end{align}
for each sampling point $P_i$ ($i=1,2,\ldots,n$). Here $\overline{x_j}$ is a random number but a constant for each set of $x_j$. Data for $y$ are computed from 
\begin{align}
	y_{i} = \beta_0 + \sum_{j=1}^{d} x_{ij} \beta_j + y_{i}'
\end{align}
in which the components of original estimator $\beta$ are known-but-withheld values, being generated randomly with the given size. The variants $x_{i1}', x_{i2}',\ldots,x_{id}', y_{i}'$ ($i=1,2,\ldots,n$), derived from Equation~(\ref{equation:variant}), are added into the estimator. Without these variants, the computed estimator is the same as the original estimator, i.e. $\hat\beta = \beta$.

For testing, we adopt the linear regression form with intercept (Equation~(\ref{equation:intercept})) as all MRs could be applied. All random numbers are bounded by the range of $[-100,100]$. In testing a mutant, a dataset is regarded as the source input.

\subsection{Assessment}

We validate the relation $\mathcal{R}$ against the sets $I_s, O_s, I_f$ and $O_f$ to determine whether the MR is violated. For instance, assume we have the source output $O_s=\{\hat\beta^s\}$ for a given source input $I_s$. If we create a follow-up input $I_f$  by adding a point generated from regression line obtained from the source output as per Equation (\ref{mr:inserting_predicted}) and reusing all original points, then we expect the follow-up output $O_f$ to have the condition $\hat\beta^s-\hat\beta^f=0$ according to Equation (\ref{mr:inserting_predicted_output}). 

Round-off errors always exist in regression computation that affects the accuracy of estimator \citep{higham2002}. As a result, we are unable to know the true values of estimator. However, we can somehow estimate the bound of error in the computed estimator. In this study, we use the first-order approximation of forward error bound $\delta$ associated with the estimator ($\hat\beta$), which is computed from the product of backward error and condition number bounds of linear regression. We adopt the backward error bound in \citet{walden1995} and \citet{chang2020}, and the condition number estimation from \citet{winkler2007}. The forward error bounds for the source ($\delta^s$) and the follow-up ($\delta^f$) are used to determine whether the computed estimators $\hat\beta^s$ and $\hat\beta^f$ are acceptable solutions of the regression program that finds the true value $\beta^s$ and $\beta^f$, respectively. As for above-mentioned example, a buggy mutant is regarded to be revealed when the source ($\hat\beta^s$) and follow-up ($\hat\beta^f$) estimators satisfy the following relationship
\begin{align}
	\norm{\hat\beta^s - \hat\beta^f}_2^2 > \norm{\delta^s}_2^2 + \norm{\delta^f}_2^2
\end{align}
that is, the relevant MR is said to be violated.

For the measurement of failure detection effectiveness of an MR, we use the metric of {\it ratio of violation} which is defined with reference to a set of mutants $\{M_1, M_2,\ldots,M_n\}$ and a set of MTGs for this MR, namely $\{\mathit{MTG}_1, \mathit{MTG}_2,\ldots,\mathit{MTG}_m\}$.
Each mutant $M_i$ is executed using every $\mathit{MTG}_j$ to see whether the MR is violated or satisfied by this pair of $(M_i,\mathit{MTG}_j)$. The ratio of violation for this MR is defined as follows
\begin{equation}
\begin{aligned}
	\text{ratio of violation} = \frac{\text{number of $(M_i,\mathit{MTG}_j)$ violating MR }}
		{\text{[ number of $(M_i,\mathit{MTG}_j)$ violating MR}}\\ +~\text{number of $(M_i,\mathit{MTG}_j)$ satisfying MR ]}
\end{aligned}
\end{equation}
In this study with the exception in Section 6, we only consider pairs of $(M_i,\mathit{MTG}_j)$ of which the outputs have the correct numeric formats in calculating the ratio of violations. Such pairs are referred to as the survived pairs of $(M_i,\mathit{MTG}_j)$.

\vspace{0.25cm}
\section{Effectiveness of Metamorphic Testing}
\vspace{0.25cm}
\subsection{Synoptic effectiveness}

\RQ{1}{Are these MRs effective in revealing failure?}

\noindent{We have carried out a total number of 9.746 $(=4430\times 100 \times 11 \times 2)$ millions of individual experimental executions. As shown in Table~\ref{table:mutation_score1}, the numbers of mutants survived after having both source and follow-up test cases with correct numeric output format range from 3127 to 3146 (which are averaged from 100 MTGs) over the total of 4430 (=9177-4747) non-equivalent mutants used. In other words, a quarter of 4430 non-equivalent mutants can be detected as faulty just by simply checking the formats of outputs.

\begin{table*}[tbp]
\begin{adjustwidth}{-0.5cm}{}
\begin{center}
{
\caption{Statistics of the number of mutants of which the outputs have correct numeric format for a given MTG (out of a total of 4430 non-equivalent mutants and over 100 samples of MTGs).}
\vspace{0.5cm}
\scalebox{0.8}{
\begin{tabular}{|r|c|c|c|c|c|c|c|c|c|c|c|}
\hline
	& \multicolumn{11}{c|}{\textbf{MR}}\\
\cline{2-12} 
\textbf{Statistics\qquad\qquad\qquad} 
	&   1.1 &  1.2  &  2.1 &   2.2 &   3.1 &   3.2 &   4.1 &   4.2 &   5.1 &  5.2 &   6  \\ 
\hline
 average   & 3127   & 3140   & 3139   & 3138   & 3135   & 3125   & 3146   & 3139   & 3140   & 3135   & 3140   \\ 
 median   & 3128   & 3130   & 3131   & 3128   & 3123   & 3116   & 3137   & 3128   & 3131   & 3126   & 3130   \\ 
 minimum   & 3067   & 3073   & 3075   & 3078   & 3063   & 3063   & 3077   & 3072   & 3074   & 3071   & 3071   \\ 
 maximum   & 3196   & 3257   & 3246   & 3239   & 3251   & 3236   & 3263   & 3258   & 3237   & 3233   & 3249   \\ 
 standard deviation   & 29   & 39   & 39   & 38   & 40   & 38   & 39   & 39   & 38   & 38   & 38   \\ 
\hline
\end{tabular}
}
\label{table:mutation_score1}
} 
\end{center}
\end{adjustwidth}
\end{table*}

We then look at how MT can help reveal failures for the remaining mutants. Table~\ref{table:mutation_score} presents the ratios of violations of all 11 MRs. It shows that all proposed 11 MRs are effective in detecting failures, having an average ratio of violations equal to 31.64\%. 

\begin{table*}[tbp]
\begin{adjustwidth}{-0.5cm}{}
{
\caption{MR's ratio of violations}.
\vspace{0.5cm}
\begin{center}
\scalebox{0.8}{
\begin{tabular}{|l|c|c|c|c|c|c|c|c|c|c|c|}
\hline
	& \multicolumn{11}{c|}{\textbf{MR}}\\
\cline{2-12} 
\textbf{Summary} 
	&   1.1 &  1.2  &  2.1 &   2.2 &   3.1 &   3.2 &   4.1 &   4.2 &   5.1 &  5.2 &   6  \\ 
\hline
Number of survived {pairs} & 312673   & 313994   & 313938   & 313804   & 313501   & 312480   & 314582   & 313914   & 314029   & 313520   & 314004   \\ 
Number of violation &  154735   & 115810   & 27260   & 66983   & 49708   & 134575   & 133330   & 150516   & 44125   & 92715   & 121803   \\ 
Ratio of violation &  49.49\%   & 36.88\%   & 8.68\%   & 21.35\%   & 15.86\%   & 43.07\%   & 42.38\%   & 47.95\%   & 14.05\%   & 29.57\%   & 38.79\%   \\ 
\hline
\end{tabular}
}
\label{table:mutation_score}
\end{center}
} 
\end{adjustwidth}
\end{table*}

\RS{1}{All MRs are effective, having an average ratio of violation of 31\%.}


\noindent 
In other words, our results show that for one MT-based test execution, there are on average approximately 31.64\% chances to reveal a failure. This is quite high, despite that we have adopted a ``conservative'' approach to consider only the survived pairs of mutants and MTGs. Even with such a conservative approach, the least effective MR still has a ratio of violation of 8.68\%. That is, this least effective MR can can reveal failures with the chance of about one out of 11 tests. The finding suggests that our 11 MRs can serve as the benchmark for testing regression systems. 

\subsection{Individual performance}

\RQ{2}{Is there any difference in performances of these MRs?}

\noindent It is found that the effectivenesses of these MRs vary significantly. In testing the selected regression programs, the most effective relation (MR1.1) has a ratio of violation that is 5.5 times higher than the lowest one (MR2.1), as shown in Table~\ref{table:mutation_score}. The best ratios are achieved by MR1.1, MR4.2, MR3.2 and MR4.1, each of which individually is able to reveal failures with a chance higher than 40\% (Table~\ref{table:mutation_score}). 

MR1.1 constructs the follow-up input by adding predicted points to the source input. The experimental data show that it has 50\% chances in revealing a failure. The runner-up is MR4.2, which has the ratio of violation of 48\% and is associated with the shifting of an independent variable by a given distance. Based on the scaling of an independent variable, the MR3.2 also achieves a ratio as high as 43\%.

\RS{2}{Some MRs are much effective than other MRs in revealing failures.}

\noindent The wide range in performance indicates that the selection of ``good'' MR is important in testing the regression system. 
To further understand why there are such differences, we examine the performance of MR over different types of bugs with the following question.

\subsection{Effectiveness for certain types of bugs}

\RQ{3}{Is MRs more effective for particular types of bugs than another?}

\noindent A regression system may consist of different types of bugs when the software engineer implements the algorithm. Therefore, it is useful to quantify the effectiveness of MRs for different types of mutation, which is linked to programming bugs in practice. For this purpose, we categorize the type of buggy mutants detected by first tallying the total numbers of survived pairs of mutant and MTG (Table~\ref{table:mutation_type0}) and then obtaining the ratio of violations for each mutation group corresponding to each MR (Table~\ref{table:mutation_type}).

\begin{table*}[tbp]
\begin{adjustwidth}{-0.5cm}{}
{
\caption{Number of {survived pairs} of mutant and MTG for each mutation group}.
\vspace{0.5cm}
\begin{center}
\scalebox{0.8}{
\begin{tabular}{|l|c|c|c|c|c|c|c|c|c|c|c|c|}
\hline
	\textbf{Mutation} &  \multicolumn{11}{c|}{\textbf{MR}} & \textbf{Total} \\
\cline{2-12} 
\textbf{group}
	&   1.1 &  1.2  &  2.1 &   2.2 &   3.1 &   3.2 &   4.1 &   4.2 &   5.1 &  5.2 &   6  & \\ 
\hline
array\_construct &     \underline{5951} &     5642 &     5661 &     5661 &     5655 &     5636 &     5648 &     5642 &     5658 &     5656 &     5659 &     62469 \\ 
array\_index &     65106 &     66295 &     66043 &     65980 &     66214 &     65951 &     \underline{66342} &     66219 &     66130 &     65917 &     66252 &     726449 \\ 
array\_swap1 &     57510 &     58054 &     58069 &     57994 &     57745 &     57676 &     \underline{58148} &     58004 &     58024 &     57956 &     57985 &     637165 \\ 
array\_swap2 &     19948 &     20075 &     20068 &     20054 &     19802 &     19863 &     \underline{20093} &     20063 &     20037 &     20032 &     20053 &     220088 \\ 
condition\_if &     550 &     548 &     550 &     \underline{552} &     547 &     549 &     549 &     550 &     549 &     \underline{552} &     \underline{552} &     6048 \\ 
condition\_index &     4276 &     4313 &     4293 &     4307 &     4319 &     4291 &     \underline{4320} &     4309 &     4314 &     4318 &     4315 &     47375 \\ 
condition\_loop &     346 &     \underline{348} &     346 &     \underline{348} &     \underline{348} &     344 &     \underline{348} &     344 &     \underline{348} &     \underline{348} &     \underline{348} &     3816 \\ 
data\_simple &     602 &     \underline{606} &     \underline{606} &     \underline{606} &     524 &     604 &     \underline{606} &     605 &     \underline{606} &     605 &     \underline{606} &     6576 \\ 
function\_parameter &     28079 &     28111 &     28308 &     28284 &     28308 &     27983 &     \underline{28314} &     28116 &     28232 &     28165 &     28031 &     309931 \\ 
function\_return &     \underline{2165} &     2098 &     2099 &     2099 &     2099 &     2099 &     2099 &     2099 &     2099 &     2099 &     2099 &     23154 \\ 
logic\_combination &     \underline{3435} &     \underline{3435} &     \underline{3435} &     \underline{3435} &     \underline{3435} &     \underline{3435} &     \underline{3435} &     \underline{3435} &     \underline{3435} &     \underline{3435} &     \underline{3435} &     37785 \\ 
logic\_comparison &     35879 &     \underline{36163} &     36089 &     36067 &     36160 &     36069 &     36155 &     36142 &     36116 &     36065 &     36156 &     397061 \\ 
logic\_not &     \underline{300} &     \underline{300} &     \underline{300} &     \underline{300} &     \underline{300} &     \underline{300} &     \underline{300} &     \underline{300} &     \underline{300} &     \underline{300} &     \underline{300} &     3300 \\ 
math\_increment &     \underline{8616} &     8510 &     8434 &     8404 &     8491 &     8441 &     8502 &     8490 &     8458 &     8406 &     8498 &     93250 \\ 
math\_initial &     8958 &     8964 &     8964 &     8966 &     8966 &     8965 &     8966 &     8963 &     \underline{8969} &     8966 &     \underline{8969} &     98616 \\ 
math\_operator &     \underline{61515} &     61040 &     61226 &     61247 &     61094 &     60844 &     61258 &     61148 &     61262 &     61223 &     61266 &     673123 \\ 
math\_values &     9437 &     9492 &     9447 &     \underline{9500} &     9494 &     9430 &     9499 &     9485 &     9492 &     9477 &     9480 &     104233 \\ 
\hline
\multicolumn{12}{l}{Maximum value of each row is \underline{underlined}.}\\
\end{tabular}
}
\label{table:mutation_type0}
\end{center}
} 
\end{adjustwidth}
\end{table*}

\begin{table*}[tbp]
\begin{adjustwidth}{-0.5cm}{}
{
\caption{Mutation group's ratios of violations.}
\vspace{0.5cm}
\begin{center}
\scalebox{0.8}{
\begin{tabular}{|l|c|c|c|c|c|c|c|c|c|c|c|c|}
\hline
	\textbf{Mutation} &  \multicolumn{11}{c|}{\textbf{MR}} & \textbf{Average} \\
\cline{2-12} 
\textbf{group}
	&   1.1 &  1.2  &  2.1 &   2.2 &   3.1 &   3.2 &   4.1 &   4.2 &   5.1 &  5.2 &   6  & \\ 
\hline
array\_construct &     \underline{72.59\%} &     65.19\% &     0.09\% &     2.72\% &     32.73\% &     58.53\% &     71.44\% &     67.07\% &     8.13\% &     10.77\% &     59.73\% &     40.81\% \\ 
array\_index &     \underline{47.58\%} &     32.01\% &     10.61\% &     27.9\% &     11.52\% &     42.78\% &     38.0\% &     45.87\% &     14.45\% &     31.27\% &     34.29\% &     30.57\% \\ 
array\_swap1 &     49.87\% &     33.33\% &     10.66\% &     26.36\% &     11.64\% &     46.33\% &     38.94\% &     \underline{50.25\%} &     18.42\% &     31.92\% &     35.83\% &     32.14\% \\ 
array\_swap2 &     45.42\% &     34.78\% &     12.54\% &     26.22\% &     10.96\% &     \underline{47.61\%} &     33.42\% &     44.68\% &     22.72\% &     32.24\% &     37.28\% &     31.62\% \\ 
condition\_if &     61.64\% &     52.01\% &     0.36\% &     25.0\% &     0.0\% &     27.87\% &     \underline{62.66\%} &     57.27\% &     3.64\% &     59.06\% &     59.6\% &     37.19\% \\ 
condition\_index &     44.6\% &     31.35\% &     16.82\% &     22.8\% &     21.93\% &     \underline{47.35\%} &     35.23\% &     46.39\% &     27.93\% &     27.74\% &     29.87\% &     32.0\% \\ 
condition\_loop &     \underline{4.91\%} &     0.57\% &     0.0\% &     0.57\% &     0.0\% &     0.0\% &     1.72\% &     0.58\% &     0.0\% &     4.31\% &     4.02\% &     1.51\% \\ 
data\_simple &     52.16\% &     34.82\% &     0.17\% &     0.17\% &     41.98\% &     52.81\% &     52.97\% &     \underline{63.47\%} &     0.33\% &     12.23\% &     14.85\% &     29.63\% \\ 
function\_parameter &     \underline{59.28\%} &     57.05\% &     0.61\% &     20.95\% &     38.68\% &     52.28\% &     59.13\% &     58.41\% &     12.66\% &     22.99\% &     57.29\% &     39.93\% \\ 
function\_return &     36.86\% &     31.32\% &     0.0\% &     17.53\% &     1.05\% &     25.87\% &     \underline{39.07\%} &     31.06\% &     1.76\% &     25.54\% &     26.44\% &     21.49\% \\ 
logic\_combination &     71.47\% &     67.25\% &     0.0\% &     27.95\% &     12.14\% &     51.67\% &     \underline{72.17\%} &     72.05\% &     7.86\% &     55.92\% &     66.32\% &     45.89\% \\ 
logic\_comparison &     \underline{43.33\%} &     31.05\% &     8.01\% &     12.57\% &     10.33\% &     34.24\% &     36.35\% &     39.54\% &     9.69\% &     24.97\% &     31.53\% &     25.6\% \\ 
logic\_not &     \underline{66.67\%} &     \underline{66.67\%} &     0.0\% &     0.0\% &     0.0\% &     \underline{66.67\%} &     \underline{66.67\%} &     \underline{66.67\%} &     0.0\% &     0.0\% &     \underline{66.67\%} &     36.36\% \\ 
math\_increment &     \underline{52.33\%} &     43.43\% &     6.84\% &     15.72\% &     19.39\% &     49.67\% &     46.53\% &     51.53\% &     18.94\% &     42.92\% &     46.61\% &     35.81\% \\ 
math\_initial &     \underline{44.24\%} &     22.79\% &     12.18\% &     10.21\% &     14.3\% &     39.34\% &     40.3\% &     43.45\% &     15.17\% &     20.58\% &     25.79\% &     26.21\% \\ 
math\_operator &     \underline{48.79\%} &     38.01\% &     9.01\% &     19.12\% &     16.85\% &     39.02\% &     45.26\% &     46.9\% &     10.02\% &     29.15\% &     40.33\% &     31.13\% \\ 
math\_values &     \underline{52.95\%} &     35.34\% &     6.03\% &     10.71\% &     19.16\% &     36.29\% &     40.54\% &     48.04\% &     12.24\% &     38.88\% &     44.91\% &     31.37\% \\ 
\hline
\multicolumn{12}{l}{Maximum percentage of each row is \underline{underlined}.}\\
\end{tabular}
}
\label{table:mutation_type}
\end{center}
} 
\end{adjustwidth}
\end{table*}

The effectiveness of MR is subject to the type of defects existed in a system. In our situation, MR1.1 can help revealing 73\% the bugs associated with the wrong construction of array (array\_construct). Their runner-ups for this type of mutation are MR4.1, MR4.2, MR1.2, and MR3.2, which have the {ratio} around 60\% or higher. MR1.1 is also able to further detect a half or more of buggy mutants associated with wrong handling of logics and branching (logic\_combination, 71\%; logic\_not, 67\%; and conditional\_if, 62\%) one-dimensional array (array\_swap1, 50\%), misuse of arguments for function and procedure (function\_parameter, 59\%), bugs in computing mathematical values (math\_values, 53\%; math\_increment, 52\%; and math\_operator, 49\%) as well as misplaced indices of array (array\_index, 48\%).  

On average, all MRs have 30\% or more chances to reveal failures in 13 out of 17 mutation groups (Table~\ref{table:mutation_type}). Note that the ratios of violation of MR for 2 mutation groups (condition\_loop and logic\_not) are not as reliable as others because they consist of a much smaller number of pairs $(M_i,\mathit{MTG}_j)$ for testing.


\RS{3}{Different MRs have different effectivenesses in revealing different types of faults.}

\subsection{Nature of data manipulation}

\RQ{4}{Is there any relationship between the effectiveness of MR and the way to construct the follow-up test cases?}

\noindent We observe that modifying the values of dependent variable may reveal less failures than changing the independent ones (Table~\ref{table:mutation_score}). Scaling the dependent variable (MR3.1) has the ratio of violation to be one third of the ratio of the squeeze of the independent variable (MR3.2) (Table~\ref{table:mutation_score}). Reflecting the dependent variable (MR2.1) also has the ratio that is half of the changing the sign of an independent one (MR2.2). Shifting the dependent variable (MR4.1) likewise has a lower ratio (by 5\%) than shifting the independent one (MR4.2). 

On the other hand, we found that the scaling of the dependent variable (MR3.2) has a higher ratio of violation than the simple change of the its sign (MR2.2), although they all originate from the same property $P2$ of estimator. In addition, swapping independent variables (MR5.2) is more effective than swapping the data points (MR5.1).



\RS{4}{MRs related to manipulating independent variables are more effective than the dependent one. Furthermore, MRs which are associated with scaling and swapping independent variables are more effective than the MRs related to reflecting variables and swapping data rows, respectively.}

\section{A comparison with random testing}

Regression systems have no test oracle (also referred to as the untestable systems). That is, for any input, we are unable to validate whether or not the computed result is correct. Nevertheless, for some special or trivial inputs, their outputs may be well known prior to computations. Thus, it is common to test an untestable system with such special or trivial test cases \citep{chen2004}. Such an approach is of limited capability because the amount of such special inputs is negligible as compared with the number of all feasible inputs. Another way that we can notice incorrect results, is by the occurrence of crashes which include improper halting of the program execution, overflow; or by having a ``long runtime'', which is significantly longer than the duration normally needed to execute a regression program, given the same input. In this study, if a mutant has the runtime that is 1000 times longer than the one of original program in all 100 test cases, it will then be regarded as having a long runtime, and hence will be considered as a failure. Irrespective of whatever method used to generate test cases, they face the same problem that incorrect results could only be revealed through the occurrence of crashes. In fact, crashes are often caused by omission of checking some conditions or omission of implementing some functions, where white-box test case generation is generally less useful because it cannot generate test cases from existing code to detect the fault related to missing code. Furthermore, current literature does not show which test case generation method is the best to lead to crashes. Therefore, we chose the simplest but also the most unbiased technique, namely, random testing (RT), as a benchmark for comparison with our method. That is, we try to answer the following research question:

%

%

\RQ{5}{Does MT perform better than RT in revealing failures?}

\noindent In doing so, we take into account the fact that linear regression systems perform many mathematical computations, and thus it is possible for a randomly generated data to cause the program under test to run into run-time computational errors such as being improperly halted, producing non-numeric outputs (i.e., `not-a-number'' (NaN)), having long runtime, or outputting wrong data size. They are the criteria to assert the effectiveness of RT in the comparison.

For a fair comparison between MT and RT, we need to resolve four issues. First, we need to use the same number of test cases for both testing methods. Since our 11 MRs use one source test case, which is randomly generated, and one follow-up test case, which is computed based on the relevant MR, we need to use two randomly generated test cases in RT when applying MT once. We denote this random testing approach as R2, with the `2' indicating that we are using two randomly generated test cases for comparison with MT.

Second, we need to deal with the issue of determining whether the program under test succeeds or fails for random testing. For testing a program that implements linear regression with randomly generated data, if the program returns a linear regression model (that is, an estimator $\hat\beta$), there is no way that we can tell whether the program succeeds or fails, due to the lack of test oracle. On the contrary, if the output of the program consists of errors such as not-a-number, empty value, null output, and incorrect dimension of estimator, we know that the program fails. Therefore, we adopt this ``working definition'' of revealing a failure for a mutant by a randomly generated test case. However, this causes another issue that we cannot use the same approach as in our experimentation with MT in Section~5 which is based on the survived pairs $(M_i,\mathit{MTG}_j)$ that have estimators in correct numeric format). As a result, we have to perform our comparison experiment using all possible (443000) pairs of mutants and test sets instead of employing only survived pairs as used in Section~5. Note that in MT, the test set is referred to as $MTG_j$ associated with a MR; while in RT, the test group consists of a pair of random datasets.

Third, we need to deal with the issue whether the difference in performance between MT and R2 is really caused by the MR and not by the ``randomly generated source test case''. To resolve this, we reuse the randomly generated source test case in MT as the first randomly generated test case in our R2. By doing this, we have a common ground for better comparison between MT and R2. The second test case in R2 will be generated randomly whereas the follow-up test cases in our MT will be generated based on each individual MR. As a result, for each source test case, we have 11 follow-up test cases, one per MR.

Fourth, we perform the following process to compare MT with R2 100 times to avoid pre-mature conclusions with just a few instances of comparison. We generate an R2 test set and 11 individual MT test sets, one per MR, using the approach discussed earlier. For each of the 4430 mutants, we execute the mutant with these test sets. We then record whether a mutant is revealed as failure by a particular test set. Each test set has only two test cases as mentioned earlier. For R2, a pair of mutant and test set can be referred to as failure if the mutant fails on any test case in the test set. For MT, the pair of mutant and MTG is regarded as failure for either one the following two reasons. One scenario is that the program runs into ``error'' with either the source test case or the follow-up test case. 
The other scenario is to have the associated MR to be violated.
We then define the {\it extended ratio of failure detection} with reference to a set of mutants and a test set for the performance comparison between of R2 and MT as follows
\begin{equation}
\begin{aligned}
	\text{extended ratio of failure detection} = \frac{\text{number of failed pairs}}
		{\text{number of pairs}}
\end{aligned}
\end{equation}
Altogether, we have a ratio for R2 and a list of 11 ratios for MT, each per MR over 100 test sets.


%

The random testing (R2) using two test cases has the {extended ratio of failure detection of 32.76\% (Figure~\ref{figure:compare}). The R2 ratio is about {two thirds} of the ratios of our MT, whose median is about {55.26\% and mean is 51.59\% {(Figure~\ref{figure:compare})}. It is half the ratio (64.34\%) of the most effective MR}. Such results allow us to assert that MT outperforms random testing in the testing scenario} in which both random testing and MT involve the same number of test cases. 

\begin{figure}[t]
	\centering
	\includegraphics[scale=0.7]{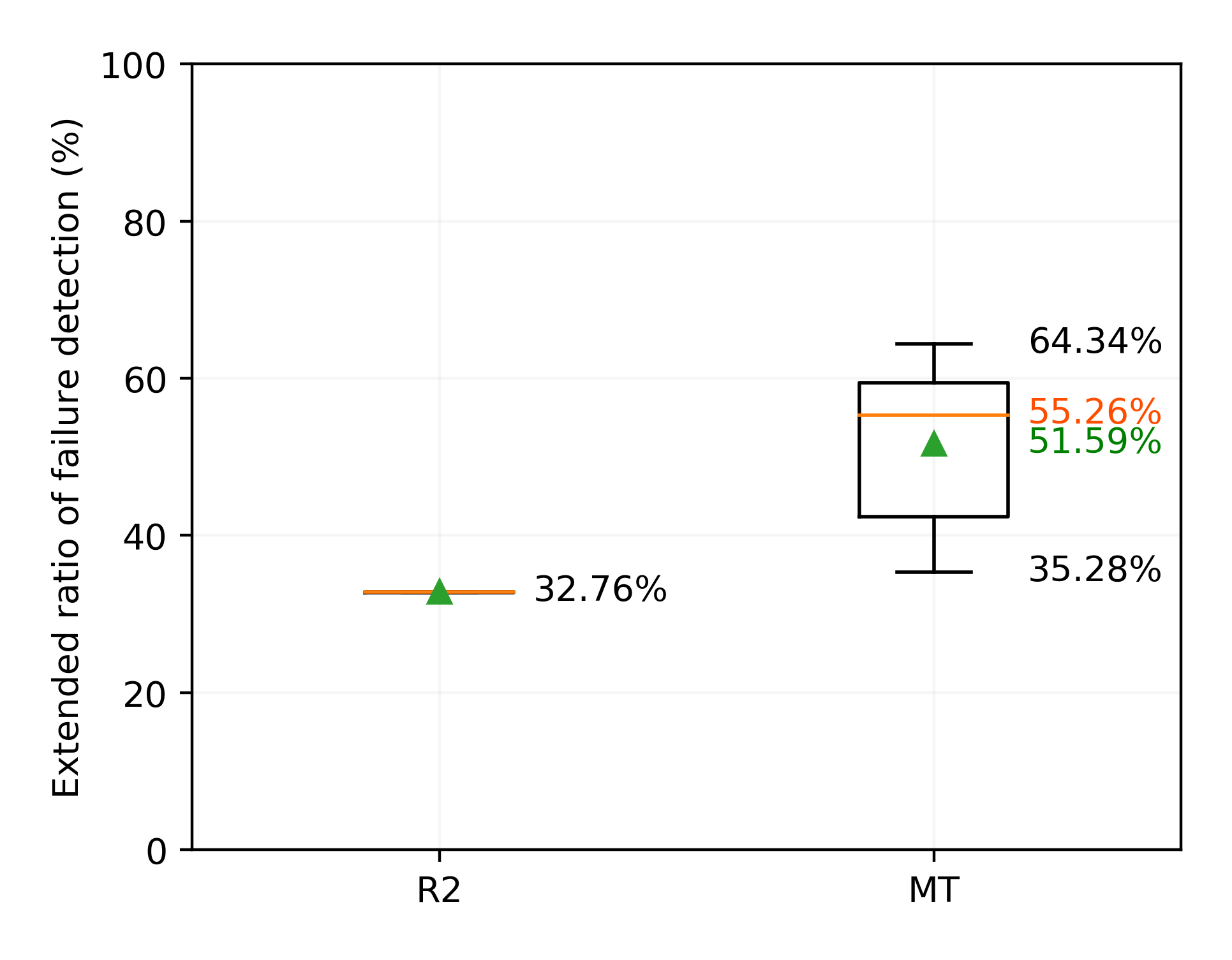}
	\caption{Comparison of effectivenesses of Metamorphic Testing (MT) against random testing (R2) using the set of 4430 mutants and 100 datasets. The percentile of MT is constructed from the ratios of failure detection of all 11 MRs. The green triangles and the green number represent the arithmetic means.}
	\label{figure:compare}
\end{figure}

\RS{5}{MT is more effective than random testing in detecting failures.}

\noindent MT outplays R2 because it utilizes the relationship between two outputs together with their inputs for validation. If we adopt more effective MRs (MR1.1, MR3.2 or MR4.2), the extended ratio of failure detection can be even higher, with the range of 59\%-64\%. Although this comparison (using all 4430 non-equivalent compilable mutants and 100 datasets) between MT and R2 gave us certain insights into the effectiveness of MT, it is worth noting that R2 is not able to cope with test oracle problem that MT is designed for.


\section{Threats to validity}

Among factors that may threaten the internal validity of our study, the main concern is about the faults of the source programs being used to conduct the experiment. Prior to the mutation analysis, the source programs have been subject to a careful manual review, whilst its complied version has been tested rigorously in both MacOS and Windows operating systems against random datasets. Together with the fact that these programs have been well tested before being published (we only modified those statements which are related to input and output functions, and such changes had been carefully checked), no fault in MR violation after a total number of 500 test cases (100 cases per program) indicates that such programs are reliable to be used in our mutation analysis. 

The second factor that may affect the causal relation in our study is the experimental bias, which is mainly attributed by the datasets selected for the testing and the uniqueness of mutants. To address this concern, we have randomized not only the regression values, but also the number of independent variables and the size of dataset, so that the impact of sampling bias is the most minimal. In addition to the comparison against the programs with 100 different datasets, we have spent a large amount of time in comparing the sources of mutant, as well as examining their semantics to be confident that each mutant used is unique. Besides, while the mutation tool allows us to arguably obtain reasonable results with a large number of non-equivalent executable mutants (4430), the ratio of violation depends on the source program and metamorphic test groups involved. The mutation keywords used (Table~\ref{table:mutation_score}) were reasonably generic, so that they are applied to generate mutants for a wide range of programs efficiently (Table~\ref{table:put_summary}). In comparison with random testing, the ``extended'' effectivenesses of MT and R2 might not be the same as Figure~\ref{figure:compare} in case a different number of test sets is used. However, a large number of test cases (4.873 millions for MT and 0.443 millions for RT) implies that the difference between MT and R2 is likely to remain significant.

\begin{figure}[t]
	\centering
	\includegraphics[scale=0.65]{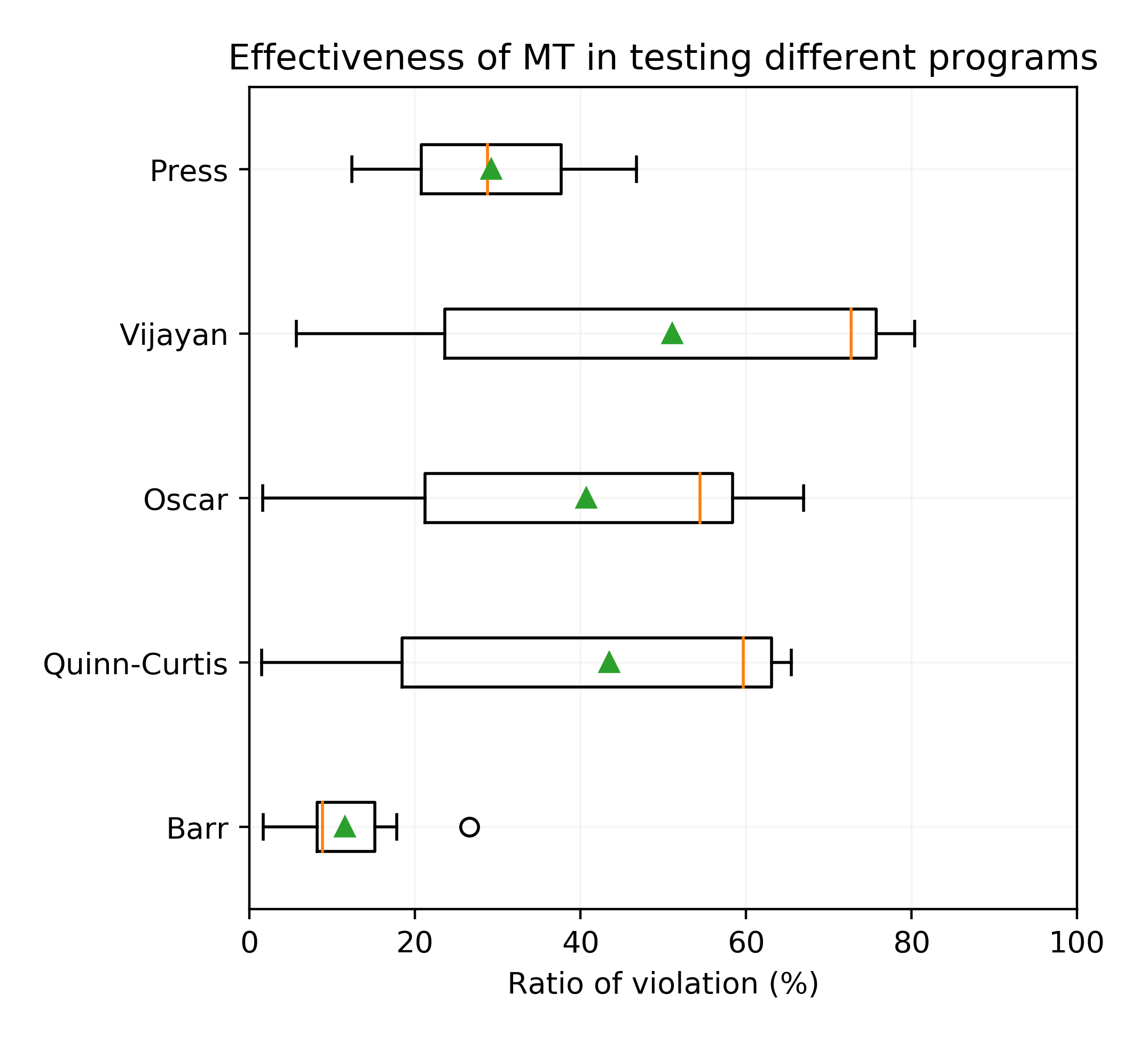}
	\caption{The effectiveness of MT in testing different programs. The green triangles represent the arithmetic means; and the open black circle denotes the outlier. The percentiles are constructed from the ratios of violation of 11 MRs.}
	\label{figure:compareputs}
\end{figure}

Another factor that may have an impact on the rigorousness of the results is the determination of MR violation. With the involvement of round-off errors and floating point arithmetic, we can only know the true values of the estimator are within a numeric interval instead of an exact value. This will lower the ratio of violations when the the forward error bound is large. However, we believe this trade-off is necessary because it helps us avoiding the type-II statistical error, i.e., the false positive assessment asserting that a non-faulty program is faulty. It is noted to mention that while large condition number may reduce the precision of forward error, the experiments with 5 different programs against 100 random datasets return no single false-positive result. In addition to the above-mentioned experiments with 100 datasets, we further extended the testing up to 1000 datasets for the program that has the largest number of mutants ({\it Press}), and no single false-positive case was identified. Such rigorous validations confirm that our estimation of forward error is reasonable. 

The main concern with the external validity is about whether our method is applicable beyond the mutation analysis. In this study, it it worth mentioning that the mutation analysis is not based on a single but a total of five different programs, in which MT are effective in all cases as shown in Table~\ref{figure:compareputs}. Furthermore, the use of mutation analysis could still give trustworthy results by generating mutants {which are} similar to real-life faults \citep{andrews2005}. Further studies with real-life applications are worthy. This will require a close collaboration with software developers.

\section{Concluding remarks}

In summary, we have proposed a novel testing approach to validate the implementation of multiple linear regression algorithm by taking advantage of the MT. The benefit of the approach is that it alleviates the problem of absence of test oracle, which is a major constraint in verifying the regression system. The originality of this paper is that we have quantified the change of estimators under different transformations from their intrinsic mathematical properties. Formally backed by the mathematical proofs, the established formulations allow us to devise the follow-up values of estimator after adding data points, scaling inputs, shifting variables, reordering data, and rotating independent variables. We then take advantage of these properties to propose 11 different MRs to test multiple linear regression systems as well as simple linear regression estimations. 

Our approach reveals a fault in Scikit-Learn. To further examine the effectiveness of ascribed MRs, we have applied the testing on a set of five different regression programs. The mutation analysis uses 4430 non-equivalent mutants. A total number of 100 datasets are generated for testing, with a total of 9.746 millions of experimental executions. We found that all developed MRs are effective in revealing failures. Some MRs are more effective than others, which help detect up to nearly half of survived pairs of $(M_i,\mathit{MTG}_j)$. The effectiveness of MR varies in accordance with the mutation type. And in general, MT is shown to be better than random testing. The 11 MRs proposed in this study can serve as a benchmark of MRs for testing new regression systems. In addition, statistical users can take advantage of the method to ensure that there is no mistake in their process manipulating the multiple linear regression.

Predictive systems based on regression are very commonly adopted in many disciplines including economics, engineering and sciences. Our future work is to apply and extend this proposed technique to validate such systems.

\section*{Acknowledgment}

The authors would like to thank Dr D.Q. Tran {(University of North Carolina at Chapel Hill)} and Dr S.H. Dau {(RMIT University)} for useful discussion.





\bibliographystyle{model4-names}\biboptions{authoryear}
\bibliography{references.bib}


\appendix

\section{Proof for the properties of estimator}

In this appendix, we present our proofs for the properties of estimators used in the paper.

\subsection{Inserting new data}

\vspace{0.25cm}
\noindent\textbf{Proposition 1. }\textit{Suppose that the dependent variable $y$ is related to a linear relationship with independent variables $x_0, x_1, \ldots, x_d$ with the estimator $\hat\beta$ derived from the least square fitting. 
Let ${\bf y}$ and ${\bf x}_0, {\bf x}_1, \ldots, {\bf x}_d$  (all $\in {\rm I\!R}^{n}$) denote the vectors of data for the variables $y$ and $x_0, x_1, \ldots, x_d$, respectively, whose matrices are expressed in Equation~(\ref{equation:denote}).
And let $x^* \in {\rm I\!R}^{d+1}$ and ${y^*} \in {\rm I\!R}$ be a data point being added into the original data set. The linear estimator $\hat\beta^*$ obtained from the new data set $ {\bf X}^* = \begin{bmatrix} {\bf X} & x^* \end{bmatrix}  \in {\rm I\!R}^{(d+1)\times (n+1)}$ and $ {\bf y}^* = \begin{bmatrix} {\bf y} & {y^*} \end{bmatrix} \in {\rm I\!R}^{n+1}$ can be derived from the following equation
\begin{align}
    \hat\beta^* 
    = \hat\beta+ G ({y^*}-x^{*T}\hat\beta)
\end{align}
where $G$ is the vector of size $d+1$ {defined by $\bf{X}$ and $x^*$ as follows}
\begin{equation}
    G = \frac
        {({\bf X}{\bf X}^T)^{-1} x^*}
        {1+x^{*T}({\bf X}{\bf X}^T)^{-1} x^*}.
\end{equation}
} 

\noindent{\bf\textit{Proof}:} 
We can determine the updated estimator $\hat\beta^*$ from the independent matrix ${\bf X^*}$ and the dependent vector ${\bf y^*}$ by definition as follows
\begin{equation}
	\hat\beta^* = ({\bf X^*}{\bf X^*}^T)^{-1} {\bf X^*}{\bf y^*}
\end{equation}
before rewriting its in the matrix form as
\begin{align}
    \hat\beta^* 
    &= 
    \left(
        \begin{bmatrix}
            {\bf X} & x^* 
        \end{bmatrix}
        \begin{bmatrix}
            {\bf X}^T \\
            x^{*T}
        \end{bmatrix}
    \right)^{-1}
    \begin{bmatrix}
        {\bf X} & x^* 
    \end{bmatrix}
    \begin{bmatrix}
        {\bf y} \\
        {y^*} 
    \end{bmatrix}
    \nonumber\\
    &= 
    \left(    
        {\bf X}{\bf X}^T + x^*x^{*T}
    \right)^{-1}
    \left(    
        {\bf X}{\bf y} + x^*{y^*}
    \right)
\end{align}

We now apply the Sherman$-$Morrison rank-1 update formula \citep{hager1989} to expand the inverse matrix
\begin{align}
    \left(    
        {\bf X}{\bf X}^T + x^*x^{*T}
    \right)^{-1}
    &= 
    ({\bf X}{\bf X}^T)^{-1} 
    \nonumber\\ 
    &-\frac
        {({\bf X}{\bf X}^T)^{-1} 
        x^*x^{*T}
        ({\bf X}{\bf X}^T)^{-1}}
        {1+x^{*T}({\bf X}{\bf X}^T)^{-1} x^*}
\end{align}
which then transforms the parameter into 
\begin{align}
    \hat\beta^* 
    &= 
    ({\bf X}{\bf X}^T)^{-1} 
    {\bf X}{\bf y} 
    +
    ({\bf X}{\bf X}^T)^{-1} 
    x^*{y^*}
    \nonumber\\
    &-
    \frac
        {({\bf X}{\bf X}^T)^{-1} 
        x^*x^{*T}
        ({\bf X}{\bf X}^T)^{-1}
        {\bf X}{\bf y}}
        {1+x^{*T}({\bf X}{\bf X}^T)^{-1} x^*}
    \nonumber\\
    &-
    \frac
        {({\bf X}{\bf X}^T)^{-1} 
        x^*x^{*T}
        ({\bf X}{\bf X}^T)^{-1}
        x^*{y^*}}
        {1+x^{*T}({\bf X}{\bf X}^T)^{-1} x^*}
\end{align}

Using the definition of $\hat\beta$ in Equation~(\ref{equation:beta}) we yield
\begin{align}
    \hat\beta^* 
    &= 
    \hat\beta
    +\nonumber\\
    &({\bf X}{\bf X}^T)^{-1} 
    x^*
    \left[
    {y^*}
    -
    \frac
        {x^{*T}\hat\beta + x^{*T}({\bf X}{\bf X}^T)^{-1} x^*{y^*}}
        {1+x^{*T}({\bf X}{\bf X}^T)^{-1} x^*}
    \right]
    \nonumber\\ 
    &=
    \hat\beta
    +
    \frac
        {({\bf X}{\bf X}^T)^{-1} x^*}
        {1+x^{*T}({\bf X}{\bf X}^T)^{-1} x^*}
     ({y^*}-x^{*T}\hat\beta)
\end{align}

\subsection{Scaling data}

\vspace{0.25cm}
\noindent\textbf{Proposition 2. }\textit{
Suppose that the values ${\bf y}^*$ of the dependent variable are scaled by a factor of $a$ with respect to the original values ${\bf y}$, and the values of an independent variables ${\bf x}^*_k$ being factorized by a factor of $b$ with respect to the original values ${\bf x}_k$, that is 
\begin{align}
{\bf y}^* &= a~ {\bf y}\\
{\bf x}^*_k &= b ~{\bf x}_k 
\end{align}
where the constants $a$ and $b$ are non-zero real numbers ($a, b \in {\rm I\!R}\backslash \{ 0 \}$). The new estimator $\hat\beta^*$ can be computed from the original estimator $\hat\beta$ as follows
\begin{equation}\label{appendix_betascale1}
\begin{aligned}
  \hat{\beta}^* = 
  \begin{bmatrix}
    a\hat\beta_{0} \\
    a\hat\beta_{1} \\
    \ldots\\
    a\hat\beta_{k-1} \\
    \frac{a}{b}\hat\beta_{k} \\
    a\hat\beta_{k+1} \\
    \ldots\\
    a\hat\beta_d
  \end{bmatrix}
\end{aligned}
\end{equation}
}

\noindent{\bf \textit{Proof}:} 
Let us assume the matrix of independent variables ${\bf X}^*$ to be computed from the matrix ${\bf X}$, such that the rows associated with an independent variable, say ${\bf x_1}$ without the loss of generosity, to be multiplied by the constant $b$, that is
\begin{equation}
\begin{aligned}
  {\bf X}^* =   
  \begin{bmatrix}
    x_{1,0} & x_{2,0} & x_{3,0} & \ldots  & x_{n,0} \\
    b x_{1,1} & b x_{2,1} & b x_{3,1} & \ldots  & b x_{n,1} \\
    \ldots & \ldots & \ldots  & \ldots & \ldots \\
    x_{1,d} & x_{2,d} & x_{3,d} & \ldots  & x_{n,d}    
  \end{bmatrix} 
\end{aligned}
\end{equation}
The product of rescaled independent and dependent variables has the elements
\begin{equation}
\begin{aligned}
  {\bf X}^* {\bf y}^* 
 =   
  \begin{bmatrix}
    a~\sum_{i=1}^{n}{x_{i,0} y_{i} } \\
    ba~ \sum_{i=1}^{n}{x_{i,1} y_{i} } \\
    \ldots \\
    a~\sum_{i=1}^{n}{x_{i,d} y_{i} } \\
  \end{bmatrix}
\end{aligned}
\end{equation}
We also have the product $ {\bf X}^* {\bf X}^{*T} \hat{\beta}^* $ using the definition of $\hat{\beta}^*$ in Equation~(\ref{appendix_betascale1}) as follows
\begin{equation}
\begin{aligned}
  {\bf X}^* {\bf X}^{*T} \hat{\beta}^* 
	 =
  \begin{bmatrix}
    a~\sum_{j=0}^{d} \hat\beta_{j}  \left( \sum_{i=1}^{n}{x_{i,0}x_{i,j}} \right) \\
    ba~ \sum_{j=0}^{d} \hat\beta_{j}  \left( \sum_{i=1}^{n}{x_{i,1}x_{i,j}} \right) \\
    \ldots\\
    a~\sum_{j=0}^{d} \hat\beta_{j}  \left( \sum_{i=1}^{n}{x_{i,d}x_{i,j}} \right) \\
  \end{bmatrix} 
\end{aligned}
\end{equation}
From the determination of $\hat\beta$, we have $ {\bf X} {\bf X}^{T} \hat{\beta} = {\bf X} {\bf y}$. Since each element of the vector in the left-hand-side of this equation should equal the corresponding element in the vector in the right-hand-side, the multiplication by either $a$ or $b \times a$ to both sides of each equation would not change the equality. Therefore, we derive the equality
\begin{equation}
\begin{aligned}
  {\bf X}^* {\bf X}^{*T} \hat{\beta}^* 
	 =
  {\bf X}^* {\bf y}^* 
\end{aligned}
\end{equation}
As the solution of least square fitting is unique, $\hat\beta^*$ is the unique rescaled estimator that we look for.

\subsection{Shifting data}

\vspace{0.25cm}
\noindent\textbf{Proposition 3. }\textit{
Suppose that the values ${\bf y}^*$ of the dependent variable are shifted by a distance of a with respect to the original values ${\bf y}$, and the values of an independent variables ${\bf x}^*_k$ being shifted by a distance of $b$ with respect to the original values ${\bf x}_k$, that is
\begin{align}
{\bf y}^* &= {\bf y} + a~{{\bf 1}^{n}}\\
{\bf x}^*_k &= {\bf x}_k + b ~{{\bf 1}^{n}}
\end{align}
where $a$ and $b$ are real constants ($b, a \in {\rm I\!R}$). The new estimator $\hat\beta^*$  can be determined from the original estimator as follows
\begin{equation} 
\begin{aligned}
  \hat{\beta}^* = 
  \begin{bmatrix}
    \hat\beta_{0} - b \hat\beta_{k} + a \\
    \hat\beta_{1} \\
    \hat\beta_{2} \\
    \ldots\\
    \hat\beta_d
  \end{bmatrix}
\end{aligned}
\end{equation}
} 

\noindent{\bf \textit{Proof}:} 
We divide the proof into two parts. In the first part, we show that the components of the updated estimator after shifting the dependent variable as in Equation~(\ref{equation:shifted-y}) is 
\begin{equation} \label{equation:shifted-beta-1a}
\begin{aligned}
  \hat{\beta}^* = 
  \begin{bmatrix}
    \hat\beta_{0} + a \\
    \hat\beta_{1} \\
    \hat\beta_{2} \\
    \ldots\\
    \hat\beta_d
  \end{bmatrix}
\end{aligned}
\end{equation}
In fact, from the equation to determine the shifted estimator $\hat\beta^*$ from the shifted dependent variable ${\bf y}^*$, we have
\begin{equation}\label{equation:shifted-beta-2}
\begin{aligned}
	{\bf X}_{int}{\bf X}^T_{int} \hat\beta^*
	= {\bf X}_{int}{\bf X}_{int}^T \hat\beta +a~{\bf X} {\bf 1}_n  
\end{aligned}	
\end{equation}
where the matrix associated with independent variables for the regression that has intercept (${\bf x_0} = {\bf 1}_n$) is expressed by
\begin{equation}
\begin{aligned}
  {\bf X}_{int} =   
  \begin{bmatrix}
    1 & 1 & 1 & \ldots  & 1 \\
    x_{1,1} & x_{2,1} & x_{3,1} & \ldots  & x_{n,1} \\
    \ldots & \ldots & \ldots  & \ldots & \ldots \\
    x_{1,d} & x_{2,d} & x_{3,d} & \ldots  & x_{n,d}    
  \end{bmatrix} 
\end{aligned}
\end{equation}
The matrix multiplications give us $d+1$ equations to determine $d+1$ components of $\hat\beta^*$. All of them require the following condition that is correct for all $x_{i,j}$ ($i=1,2,\ldots,n$ and $j=1,2,\ldots,d$)
\begin{equation}
\begin{aligned}
\hat\beta_{0}^* + \sum_{j=1}^{d} \hat\beta_{j}^* x_{i,j}
= \hat\beta_{0} + \sum_{j=1}^{d} \hat\beta_{j} x_{i,j} + a
\end{aligned}
\end{equation}
which is equivalent to
\begin{equation}
\begin{aligned}
(\hat\beta_{0}^* - \hat\beta_{0} - a) + \sum_{j=1}^{d}x_{i,j} (\hat\beta_{j}^*-\hat\beta_{j})  = 0
\end{aligned}
\end{equation}
The only $\hat\beta^*$ to satisfy this condition for all $x_{i,j}$ is the estimator given in Equation~(\ref{equation:shifted-beta-1a}).

In the second part, we show that the components of the updated estimator after shifting the dependent variable as in Equation~(\ref{equation:shifted-x}) is 
\begin{equation}\label{equation:shifted-beta-3}
\begin{aligned}
  \hat{\beta}^* = 
  \begin{bmatrix}
    \hat\beta_{0} - b \hat\beta_{k} \\
    \hat\beta_{1} \\
    \hat\beta_{2} \\
    \ldots\\
    \hat\beta_d
  \end{bmatrix}
\end{aligned}
\end{equation}
As the matter of fact, the shifted estimator $\hat\beta^*$ is determinable from the matrix associated with the shifted independent variables ${\bf X}^*$ by definition as followings
\begin{equation}
\begin{aligned}
	{\bf X}^*_{int} {\bf X}^{*T}_{int} \hat\beta^*
	= {\bf X}^*_{int} {\bf y}
\end{aligned}	
\end{equation}
Without the loss of generosity, let us assume the independent variable ${\bf x_1}$ is shifted by the distance $b$. The matrix associated with the independent variables is then expressed by
\begin{equation}
\begin{aligned}
  {\bf X}^*_{int} =   
  \begin{bmatrix}
    1 & 1 & 1 & \ldots  & 1 \\
    x_{1,1}+b&  x_{2,1}+b & x_{3,1}+b & \ldots  & x_{n,1}+b \\
    x_{1,2} & x_{2,2} & x_{3,2} & \ldots  & x_{n,2} \\   
    \ldots & \ldots & \ldots  & \ldots & \ldots \\
    x_{1,d} & x_{2,d} & x_{3,d} & \ldots  & x_{n,d}    
  \end{bmatrix}
\end{aligned}
\end{equation}
Since $\hat\beta$ is the solution of original regression model, we have
\begin{equation}
\begin{aligned}
	{\bf X}_{int} {\bf X}^{T}_{int} \hat\beta
	= {\bf X}_{int} {\bf y}
\end{aligned}	
\end{equation}
and thus can derive the equation that links $\hat\beta^*$ and $\hat\beta$
\begin{equation}
\begin{aligned}
&{\bf X}_{int} {\bf X}^{T}_{int} 
		(\hat\beta^* - \hat\beta)
	= 
    b
  \begin{bmatrix}
    0\\
    \sum_{i=1}^{n} y_{i} \\
    0\\
    \ldots \\
    0\\
  \end{bmatrix} 	
	  \\&-b
   \begin{bmatrix}
    n~\hat\beta_{0}^*
    \\
    	 n\hat\beta_{0}^*  
    	 	+ \hat\beta_{1} \sum_{i=1}^{n}{x_{i,1}}
    	 	+ n\times b	 \hat\beta_{1}^*  
    		+ \sum_{j=1}^{n} \hat\beta_{j}^* \sum_{i=1}^{n} x_{i,j}     	
    		\\
    \hat\beta_{d-1}^* \sum_{i=1}^{n}{x_{i,d-1}} \\
    \ldots\\
    \hat\beta_{d}^* \sum_{i=1}^{n}{x_{i,d}} \\
  \end{bmatrix}
\end{aligned}	
\end{equation}
One intrinsic property of our ordinary linear regression is that it is unbiased, that is
\begin{equation}
\begin{aligned}
 	\sum_{i=1}^{n} y_{i} 
 		= \sum_{i=1}^{n} \hat{y}_{i}  
 		= n \hat\beta_{0}  +\sum_{i=1}^{n} \sum_{j=1}^{d}  \hat\beta_{j} x_{i,j}  
\end{aligned}   
\end{equation}
As a result, the criteria for $\hat\beta^*$ in relation to $\hat\beta$ is
\begin{equation}
\begin{aligned}
   	\hat\beta_{0}^* - \hat\beta_{0} + b \hat\beta_{1}^*
   		+ \sum_{j=1}^{d} (\hat\beta_{j}^* - \hat\beta_{j}) x_{i,j}  
   		= 0
\end{aligned}   
\end{equation}
for all $x_{i,j}$ ($i=1,2,\ldots,n$ and $j=1,2,\ldots,d$). The only $\hat\beta^*$ to satisfy this condition for arbitrary $x_{i,j}$ is the estimator given in Equation~(\ref{equation:shifted-beta-3}).

\subsection{Permuting data}

\noindent\textbf{Proposition 4. }\textit{
Suppose that the samples ${\bf y}^*$ of the dependent variable are permuted by the function $\sigma_{v}$ with respect to the original variable ${\bf y}$; and the samples ${\bf x}^*_0,{\bf x}^*_1,\ldots,{\bf x}^*_d$ of the independent variables are permuted by both functions $\sigma_{s}$ and $\sigma_{v}$ with respect to the original values $\{{\bf x}_0,{\bf x}_1,\ldots,{\bf x}_d\}$, such that
\begin{align}
	{\bf y}^* &= \sigma_s({\bf y})\\
	{\bf x}^*_k &= \sigma_s({\bf x}_{\sigma_v(k)})
\end{align}
where $\sigma_{v}$ is a permutation (bijective function) from set $\{{\bf x}_0,{\bf x}_1,\ldots,{\bf x}_d\}$ to $\{{\bf x}^*_0,{\bf x}^*_1,\ldots,{\bf x}^*_d\}$; whilst $\sigma_{s}$ is the corresponding bijective renumbering of the set of sample index $\{1,2,\ldots,n\}$. 
The new estimator $\hat\beta^*$ can be determined from the original estimator as follows
\begin{equation} \label{equation:appendix-permute1}
\begin{aligned}
  \hat{\beta}^* = 
  \begin{bmatrix}
    \hat\beta_{\sigma_v(0)} \\
    \hat\beta_{\sigma_v(1)} \\
    \ldots\\
    \hat\beta_{\sigma_v(d)}
  \end{bmatrix}
\end{aligned}
\end{equation}
} 
 
\noindent{\bf \textit{Proof}:} 
We have the equation determining $\hat\beta^*$ as follow
\begin{align}
	\hat\beta^*
		&= \underset{\beta}{\mathrm{arg~min}} \norm{ {\bf y}^* - \hat{\bf y}^*  } _2^2\\
		&= \underset{\beta}{\mathrm{arg~min}} \norm{ \sigma_s({\bf y}) - \sum_{k=0}^d  \beta^*_k\sigma_s({\bf x}_{\sigma_v(k)})  } _2^2
\end{align}
The permutation does not change the arithmetic sum of the quantity in the norm, since both $\sigma_{s}$ and $\sigma_{s}$ are the bijective renumbering of the set of indices $\{1,2,\ldots,n\}$. Denote $j={\sigma_v(k)}$, we have
\begin{align}\label{equation:appendix-permute2}
	\hat\beta^*
		&= \underset{\beta}{\mathrm{arg~min}} \norm{ {\bf y} - \sum_{j=0}^d  \beta^*_{\sigma^{-1}_v(j)} {\bf x}_j } _2^2
\end{align}
The solution is unique due to the fact that our OLS function is strictly convex. Therefore, Equation~(\ref{equation:appendix-permute2}) gives
\begin{align}\label{equation:appendix-permute3}
	  \hat\beta^*_{\sigma^{-1}_v(j)} = \hat\beta_j  
\end{align}
The bijective mapping allow us to deduce Equation~(\ref{equation:appendix-permute1}) from Equation~(\ref{equation:appendix-permute3}).

\subsection{Rotating data}

\vspace{0.25cm}
\noindent\textbf{Proposition 5. }\textit{
Suppose that the values ${\bf y}^*$ of the dependent variable is kept unchanged, while the components ${\bf x}^*_0,{\bf x}^*_1,\ldots,{\bf x}^*_d$ of the independent variables are rotated by the matrix ${\bf R}$ of size $(d+1)\times (d+1)$. In other words, ${\bf R}$ rotates the matrix ${\bf X}^*$ with respect to the original matrix ${\bf X}$, that is
\begin{equation}
	{\bf X}^* = {\bf R} {\bf X} 
\end{equation} 
Then the new estimator $\hat\beta^*$ can be determined from the original estimator as follows
\begin{equation}
	\hat\beta^* = {\bf R}\hat\beta
\end{equation} 
}

\noindent
{\bf \textit{Proof}:} By definition, the updated estimator $\hat\beta^*$ is determinable from the independent matrix ${\bf X^*}$ and the dependent vector ${\bf y}$ as follows
\begin{equation}  \label{equation:rotated-beta-2}
	\hat\beta^* = ({\bf X^*}{\bf X^*}^T)^{-1} {\bf X^*}{\bf y}
\end{equation}
The transpose of the product of two matrices ${\bf R}$ and ${\bf X}$ is $({\bf R} {\bf X})^T={{\bf X}^T \bf R}^T$. Taking it into account in substituting the definition (\ref{equation:rotated-beta-a}) into Equation~(\ref{equation:rotated-beta-2}), we have
\begin{equation}   \label{equation:rotated-beta-3}
	\hat\beta^* = ( {\bf R} {\bf X}{\bf X}^T {\bf R}^T)^{-1} {\bf R}{\bf X}{\bf y}
\end{equation} 
Since  ${\bf R}$, ${\bf X}{\bf X}^T$ and ${\bf R}^T$ are square matrices having the same size, the inverse of their product can be decomposed into the product of the inverse matrices in the reversed order, that is
\begin{equation}  \label{equation:rotated-beta-4}
	\hat\beta^* = 
	({\bf R}^T)^{-1} ({\bf X}{\bf X}^T)^{-1}  {\bf R}^{-1}
		 {\bf R}{\bf X}{\bf y}
\end{equation} 
One noticeable property of the {rotation} matrix is that the inverse of the given matrix can be determined from its transpose, i.e. ${\bf R}^{-1} ={\bf R}^{T}$. As a result, we have $({\bf R}^T)^{-1} = {\bf R}$. In addition, the product of the inverse matrix ${\bf R}^{-1}$ and ${\bf R}$ is equal to an identity matrix, having same size as ${\bf X}$. Therefore, we can rewrite Equation~(\ref{equation:rotated-beta-4}) as follows
\begin{equation} 
	\hat\beta^* = 
	{\bf R}({\bf X}{\bf X}^T)^{-1}{\bf X}{\bf y}
\end{equation} 
We complete proving the proposition by substituting the definition of $\hat\beta$ in this equation.

\end{document}